\documentclass[final,1p,times]{elsarticle}

\usepackage{amssymb}

\usepackage{hyperref}
\usepackage{amsmath,amsthm}

\usepackage{longtable} 
\usepackage{lscape} 

\newcounter{bla}

\journal{Computer Physics Communications}

\begin{document}

\sloppy  

\begin{frontmatter}



\title{\emph{monteswitch}: A package for evaluating solid--solid free energy differences via lattice-switch Monte Carlo}


\author[tlu]{T. L. Underwood\corref{cor1}}
\address[tlu]{Department of Physics, University of Bath, Bath, BA2 7AY, UK}
\cortext[cor1]{Corresponding author}
\author[gja]{G. J. Ackland}
\address[gja]{School of Physics and Astronomy, SUPA, The University of Edinburgh, Edinburgh,
EH9 3JZ, UK }

\begin{abstract}
Lattice-switch Monte Carlo (LSMC) is a method for evaluating the free energy between two given solid phases.
LSMC is a general method, being applicable to a wide range of problems and interatomic potentials.
Furthermore it is extremely efficient, ostensibly more efficient than other existing general methods.
Here we introduce a package, \emph{monteswitch}, which can be used to perform LSMC simulations.
The package can be used to evaluate the free energy differences between pairs of solid phases, including multicomponent phases, 
via LSMC for atomic (i.e., non-molecular) systems in the NVT and NPT ensembles. It could also be used to evaluate the free energy cost 
associated with interfaces and defects.
Regarding interatomic potentials, \emph{monteswitch} currently supports various commonly-used pair potentials, including the
hard-sphere, Lennard-Jones, and Morse potentials, as well as the embedded atom model.
However the main strength of the package is its versatility: it is designed so that users can easily implement their own potentials.
\end{abstract}


\end{frontmatter}



{\bf PROGRAM SUMMARY}

\begin{small}
\noindent
{\em Manuscript Title:} monteswitch: A package for evaluating solid--solid free energy differences via lattice-switch Monte Carlo  \\
{\em Authors:} T. L. Underwood, G. J. Ackland \\
{\em Program Title:} monteswitch                                \\
{\em Journal Reference:}                                      \\
{\em Catalogue identifier:}                                   \\
{\em Licensing provisions:}  MIT                               \\
{\em Programming language:} Fortran 95, MPI                    \\
{\em Computer:} Any                                            \\
{\em Operating system:} Unix-like                                    \\
{\em RAM:} Depends on the nature of the problem.                  \\
{\em Number of processors used:} Any                           
{\em Classification:} 7.7 Other Condensed Matter inc. Simulation of Liquids and Solids  \\
{\em External routines/libraries:}  To perform parallel simulations MPI is required (but MPI is not required to perform serial simulations). \\
{\em Nature of problem:} Calculating the free energy difference between two solid systems. \\
{\em Solution method:} Lattice-switch Monte Carlo (LSMC) [1] is a versatile and efficient method for evaluating the free energy difference between
two solid phases. The package presented here allows LSMC simulations to be performed for a variety of interatomic potentials, including commonly-used
pair potentials and the embedded atom model. Furthermore the package is designed so that users can easily implement their own potentials. 
The package supports LSMC simulations in the NVT and NPT ensembles, and can treat multicomponent systems. A version of the main program is included
which is parallelised using MPI. This program parallelises the LSMC calculation by simulating multiple replicas of the system in parallel. \\
{\em Restrictions:} monteswitch cannot treat molecular systems, i.e., systems in which the particles exhibit rotational degrees of
freedom, and is restricted to systems which can be represented within an orthorhombic supercell.
Furthermore, the interatomic potential is `hard-coded' in the sense that implementing a different potential requires that
the package be recompiled. \\
{\em Additional comments:} monteswitch includes programs to assist with the creation of input files and the
post-processing of output files created by the main Monte Carlo programs. A user manual, a suite of test cases, a worked example, and
a collection of plug-ins to implement various commonly-used interatomic potentials are also included with the package. \\
{\em Running time:} Depends on the nature of the problem and the underlying computing platform. 
For the Zr EAM example in the manuscript one iteration (i.e., one 160,000-sweep weight-function-generation simulation 
and one 700,000-sweep production simulation) took a wall-clock time of approximately 1.9 hours on a desktop machine (an iMac14,2 with a 3.2GHz Intel 
Core i5-4570 processor) exploiting 4 cores for the 384-atom system, and 17.7 hours for the 1296-atom system. 
For each ensemble in the hard-sphere example the 18,000,000-sweep weight-function-generation simulation and two 125,000,000-sweep
production simulations took a total of approximately 11 hours exploiting 16 cores on one node of a HPC cluster.
\\

\end{small}



\section{Introduction}\label{sec:intro}
The stable phase under given conditions is that with the lowest free energy. For this reason, efficiently calculating free energies is one 
of the most fundamental problems in theoretical materials science. A plethora of different methods have been developed to this end, 
each designed with a particular problem in mind (see, e.g., Ref. \cite{book:Frenkel}). Unfortunately however, commonly-used methods for 
calculating free energies of solid phases often cannot achieve the accuracy required for practical applications: an intractable amount of computational 
effort would be required. This problem is by no means limited to `complicated' models of particle interactions, but 
persists even when simple models are used. For instance it was only relatively recently demonstrated that the fcc phase is
favoured over the hcp phase in the hard-sphere solid -- an archetype of a simple system \cite{Bruce_1997,Bruce_2000,Jackson_2002}.

\emph{Lattice-switch Monte Carlo} (LSMC) \cite{Bruce_1997,Bruce_2000} 
\footnote{The reader should be aware that LSMC has also been referred to as lattice-\emph{switching} Monte Carlo.}
is a method which can be used to efficiently evaluate the free energy difference between two solid phases.
It has been applied to a wide range of systems \cite{Bruce_1997,Pronk_1999,Mau_1999,Bruce_2000,Jackson_2002,Jackson_2007,Yang_2008,
Marechal_2008,Raiteri_2010,Wilms_2012,Quigley_2014,Bridgwater_2014,Underwood_2015},
beginning with the hard-sphere solid \cite{Bruce_1997,Pronk_1999,Mau_1999,Bruce_2000}, where it was used to resolve the aforementioned hcp--fcc problem 
\cite{Bruce_1997,Bruce_2000}. The method was later applied to soft interatomic potentials \cite{Jackson_2002,Wilms_2012,Underwood_2015}, 
systems containing multiple particle species \cite{Jackson_2007,Yang_2008}, and molecular systems 
\cite{Marechal_2008,Raiteri_2010,Quigley_2014,Bridgwater_2014}.
LSMC has also inspired \emph{phase switch Monte Carlo}, a method for calculating the free energy difference between a solid and a
fluid phase \cite{Wilding_2000}, which has also seen some use
\cite{Wilding_2000,Errington_2004,McNeil-Watson_2006,Wilding_2009_MP,Wilding_2009_JCP,Sollich_2010,Wilding_2010}.
As well as being versatile, LSMC is an accurate method: it is `exact' in the sense that it relies upon no approximations other than those present in 
the model of particle interactions it is used in conjunction with. Moreover for the purposes of evaluating the free energy difference between 
pairs of solid phases LSMC is ostensibly more efficient than other existing general methods \cite{Wilms_2012,Marechal_2008}.
\footnote{We do not include methods rooted in the harmonic approximation (including the quasi-harmonic approximation \cite{Fultz_2010,vandeWalle_2002}) 
within the class of `general methods' mentioned here: these methods are not `general' in the sense that they break down in the anharmonic regime.}
\footnote{To elaborate, in Refs. \cite{Wilms_2012,Marechal_2008} LSMC was shown to significantly outperform thermodynamic integration (TI)
\cite{book:Frenkel,Vega_2008}. However the claim that LSMC outperforms TI has proved contentious \cite{Pronk_1999,Sweatman_2015}.
Of course, like-for-like comparisons between the two methods are difficult, since different implementations of LSMC or TI may be more or less
efficient than other implementations. We believe that the claim that LSMC is \emph{at least} as efficient as TI reflects the findings of 
studies up to the present time.}
However, despite its strengths, LSMC has unfortunately yet to have gained widespread popularity. This stems in part from the lack of 
an LSMC code which is both widely available and applicable to a wide range of systems.

With this in mind we have developed a package, \emph{monteswitch}, which implements the LSMC method. The package, written in Fortran 95, can 
be used to evaluate free energy differences between pairs of solid phases in the NVT and NPT ensembles. Furthermore the
package contains a version of the main executable which is parallelised using MPI for HPC applications.
Note that the two `phases' under consideration need not necessarily be homogeneous crystals; 
an interesting prospect is to use \emph{monteswitch} to evaluate free energy costs associated with interfaces and defects -- the former is something which
has been done previously using LSMC \cite{Pronk_1999}.
\footnote{We describe how LSMC can be used to evaluate interfacial free energies in Section \ref{sec:conclusions}.}
Furthermore \emph{monteswitch} can treat systems containing multiple species of particles.
However it should be noted that \emph{monteswitch} can only treat `atomic' systems (i.e., `non-molecular' systems: those in which the constituent 
particles do not have rotational degrees of freedom), and pairs of phases which can be represented by orthorhombic unit cells.

While steps have been recently been taken to implement LSMC in an existing general-purpose code,
\footnote{Specifically, LSMC is earmarked for inclusion in the general-purpose Monte Carlo code \emph{DL\_MONTE} \cite{Purton_2013}.}
we believe that \emph{monteswitch} will fulfil an important `gap in the market' for the foreseeable future because it was designed from 
the outset to be highly-customisable with regards to the interatomic potential. By contrast general-purpose codes tend to have a fixed set 
of interatomic potentials to draw upon.
In \emph{monteswitch} all of the procedures pertaining to the interatomic potential are housed within a single Fortran module. 
It is intended that users write their own version of this module which implements the interatomic potential they are 
interested in.
\footnote{Of course, in doing this the user's module is free to interface with `external' modules, or even external programs.}
(A similar scheme is utilised in the molecular dynamics program MOLDY \cite{Ackland_2011}). 
Templates are provided with \emph{monteswitch} to assist with this. Furthermore modules are included
with \emph{monteswitch} which correspond to some commonly-used interatomic potentials, which can serve as examples. Of course these modules can also be
used within \emph{monteswitch} to perform LSMC calculations.

Here we provide an introduction to \emph{monteswitch}. Note that much of what follows is elaborated upon in \emph{monteswitch}'s user manual (included
with the package), where we direct interested readers for more details. The layout of this work is as follows. In the next section we describe the
theory which underpins \emph{monteswitch}. In Section \ref{sec:preliminaries} we provide an overview of what is included in the 
\emph{monteswitch} package. In Section \ref{sec:potentials} we describe how interatomic potentials are implemented in \emph{monteswitch}, list the
various interatomic potentials included with \emph{monteswitch}, and describe how users can implement their own potentials. 
In Section \ref{sec:simulation_programs} we describe the main Monte Carlo programs within \emph{monteswitch}. 
In Section \ref{sec:utility_programs} we describe the various utility programs included with \emph{monteswitch} 
for the creation of input files and post-processing of output files.
In Section \ref{sec:example} we provide two examples to elucidate how \emph{monteswitch} could be used in practice. In the first example we apply
\emph{monteswitch} to the hard-sphere solid, and test \emph{monteswitch} against known LSMC results for this system; in the second we use 
\emph{monteswitch} to determine the hcp--bcc transition temperature and related quantities for an embedded atom model of Zr.
Finally in Section \ref{sec:conclusions} we present our conclusions and outlook.

\section{Theoretical background}\label{sec:LSMC}

\subsection{Calculating free energy differences}
Consider a system which is free to visit two phases 1 and 2 (and only phases 1 and 2).
The equilibrium phase is that with the lower free energy $\mathcal{F}$, where $\mathcal{F}$ is the Helmholtz free energy in the 
NVT ensemble and the Gibbs free energy in the NPT ensemble. It is the free energy difference between the phases 
$\Delta\mathcal{F}\equiv \mathcal{F}_1-\mathcal{F}_2$ which we wish to evaluate, where $\mathcal{F}_1$ and $\mathcal{F}_2$ denote the free energies of 
phases 1 and 2.
It can be shown that
\begin{equation}\label{DeltaF_stat_mech}
\Delta \mathcal{F}=\beta^{-1}\ln\biggl(\frac{p_2}{p_1}\biggr),
\end{equation}
where $p_1$ and $p_2$ denote the probability of the system being in phase 1 and 2 respectively, $\beta=1/(k_BT)$, $k_B$ denotes Boltzmann's constant, and
$T$ denotes the temperature of the system.
For a simulation which samples the ensemble under consideration, e.g., molecular dynamics, $p_2/p_1$ can be determined: measure the relative time $t_1$ 
and $t_2$ which the system spends in each phase 1 and 2 during the simulation, and substitute these quantities into the above equation, bearing in mind 
that $t_2/t_1=p_2/p_1$ for a sufficiently long simulation.Hence $\Delta\mathcal{F}$ can in principle be obtained from such a simulation via the above equation.
However, this method is usually intractable in practice for two solid phases, because the time taken for the system to transition between the two phases 
is too long to allow a reasonable estimate of $p_2/p_1$ to be deduced in a reasonable simulation time. It may even be the case that, regardless of 
the phase in which the simulation is initialised, the system \emph{never} transitions to the `other' phase during the course of the simulation.
The problem is that, while the regions of phase space corresponding to phase 1 and phase 2 both correspond to probable states of the system
at thermodynamic equilibrium, these regions are separated by a \emph{free energy barrier} -- a region of phase space associated
with states which are very improbable at thermodynamic equilibrium. This barrier inhibits transitions between the regions of phase space associated 
with phase 1 and phase 2.

This problem can in principle be circumvented with the Monte Carlo method. In the original incarnation of Monte Carlo, which we refer to 
as \emph{canonical Monte Carlo}\cite{Metropolis_1953} (which we contrast with \emph{multicanonical} Monte Carlo later), the system is evolved during the 
simulation as follows. Each time step we generate a trial state of the system $\sigma'$, and attempt to change the system to the trial state
from its current state $\sigma$.
The traditional approach for NVT ensembles is to perform a `particle move' to generate a trial state. Here, one particle in $\sigma$ is moved
to yield $\sigma'$. In NPT ensembles particle moves are supplemented by `volume moves', in which the volume, and potentially the shape, of the
entire system is altered, along with a commensurate rescaling of the particle positions. We accept the change of state from $\sigma$ to $\sigma'$ with
a probability $p_{\sigma\to\sigma'}$, which is a function of the energies of the states $\sigma$ and $\sigma'$ in the NVT ensemble, and
the enthalpies and volumes of the states $\sigma$ and $\sigma'$ in the NPT ensemble. The function also depends on the specific scheme used to generate
state $\sigma'$ from $\sigma$ (see, e.g., Ref. \cite{book:Frenkel}). 
The end result is that each state $\sigma$ is sampled with a probability $p_{\sigma}$ which reflects the underlying ensemble, e.g., for the NVT ensemble:
\begin{equation}
p_{\sigma}\propto e^{-\beta E_{\sigma}},
\end{equation}
where $E_{\sigma}$ denotes the energy of state $\sigma$.
However, in canonical Monte Carlo one has considerable freedom as to how trial states are generated; one is by no means limited to the aforementioned
`traditional' move set. The prospect therefore exists of generating trial states in a manner which results in the system traversing a path in phase
space which allows $\Delta \mathcal{F}$ to be calculated in a reasonable simulation time. Such a path would involve frequent transitions between both 
phases 1 and 2 by `tunnelling' through the free energy barrier separating them.

\subsection{Lattice-switch moves}
\begin{figure}
\centering
\includegraphics[width=0.9\textwidth]{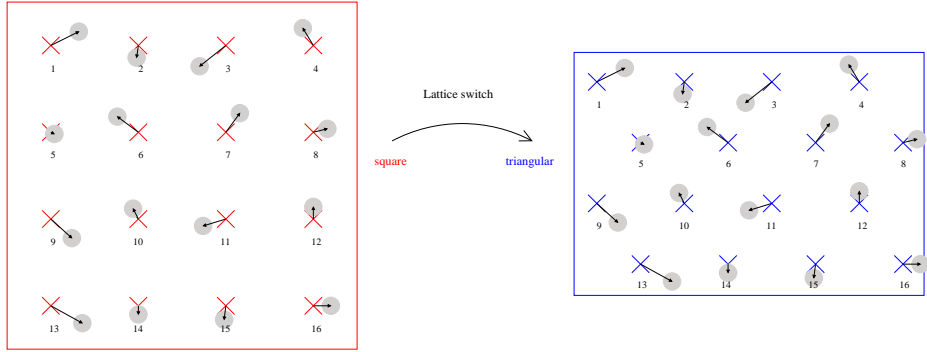}
\caption
{Schematic diagram illustrating a lattice switch from a state in the square phase of a notional two-dimensional system to a state in the triangular phase. 
In the lattice switch the underlying square lattice (red crosses) is transformed into a triangular lattice (blue crosses), while the displacements of 
the particles from their lattice sites (black arrows) is unchanged -- the displacement for each particle $n$ is the same before and after the lattice 
switch. Note also that the lattice switch here transforms the shape of the system: the red box is transformed into the blue box.}
\label{fig:switch_diagram}
\end{figure}

In LSMC a new type of move, a \emph{lattice switch}, is introduced to supplement the traditional move set mentioned above. A lattice switch move takes
the system \emph{directly} from one phase to the other, bypassing any free energy barriers separating the phases. Every time a lattice switch is accepted, 
the system transitions to the `other' phase. The salient feature of the move is that the underlying `lattice' which characterises the current phase is 
`switched' for a lattice which characterises the other phase, while the \emph{displacements} of all particles from their associated lattice sites are
preserved. This is illustrated in Fig. \ref{fig:switch_diagram} for the square and triangular phases of a notional two-dimensional system.

More formally, we can characterise a given state of the system as belonging to a solid phase $\alpha$ if the positions of the particles 
`approximately' form a lattice characteristic of $\alpha$. Let $\lbrace\mathbf{R}^{(\alpha)}_i\rbrace$ denote the positions of the sites on this lattice, 
and let $\lbrace\mathbf{r}_i\rbrace$ denote the positions of the particles, where $i$ ranges from 1 to the number of particles in the system. 
The position $\mathbf{r}_i$ of particle $i$ can be expressed as follows:
\begin{equation}
\mathbf{r}_i=\mathbf{R}^{(\alpha)}_i+\mathbf{u}_i,
\end{equation}
where $\mathbf{u}_i$ is the displacement of $i$ from that lattice site. Note that the displacements $\lbrace\mathbf{u}_i\rbrace$ are necessarily small 
since the particle positions form an approximate $\alpha$ lattice (and we have chosen to label particles and lattice sites in a `sensible' manner: such 
that $\mathbf{R}_i$ is the closest lattice site to $\mathbf{r}_i$). Now, in a lattice switch from phase $\alpha$ to the `other' phase $\alpha'$ we 
transform the underlying lattice $\lbrace\mathbf{R}^{(\alpha)}_i\rbrace$ to $\lbrace\mathbf{R}^{(\alpha')}_i\rbrace$, 
\emph{while keeping the particle displacements $\lbrace\mathbf{u}_i\rbrace$ unchanged}. The result is that the trial state belongs to phase 
$\alpha'$: the positions of the particles in the trial state form an approximate $\alpha'$ lattice.

Of course, the above description of a lattice switch is not a complete account of how lattice switches are implemented in \emph{monteswitch} -- which
supports lattice switches which change the shape and size of the supercell, as well as the species of the particle. Details of how lattice switches
are implemented in \emph{monteswitch} can be found in the user manual.

\subsection{Multicanonical Monte Carlo}
One might expect that by regularly making lattice switches, the system will regularly transition between phases, and hence
allow $\Delta \mathcal{F}$ to be efficiently evaluated as described above.
Unfortunately using canonical Monte Carlo one finds that lattice switches are too rarely accepted for this approach to be useful. 
The problem is that the trial state $\sigma'$ generated by a lattice switch is almost always of much higher energy than the current state $\sigma$, 
and hence will almost always be rejected.
\footnote{The situation is slightly more complicated for lattice switches which change the system volume. In this case the extent to which the
volume of the system is expanded/contracted influences how likely the lattice switch is to be successful. Accordingly the order parameter defined 
later in Eqn. \eqref{M_def} for `selecting' gateway states (defined in a moment) takes a slightly different form for volume-altering lattice switches
in \emph{monteswitch} -- see the \emph{monteswitch} user manual for more details. Aside from that the forthcoming discussion applies generally: to
both volume-altering and volume-preserving lattice switches.}
The solution to this problem is to use \emph{multicanonical Monte Carlo}\cite{Berg_1991,Berg_1992,Smith_1995} instead of canonical Monte Carlo. 
Multicanonical Monte Carlo can be regarded as canonical Monte Carlo, but if the energy for each state $\sigma $ were 
\begin{equation}
\tilde{E}_{\sigma}=E_{\sigma}-\eta_{\sigma}/\beta
\end{equation}
instead of $E_{\sigma}$, where $\eta_{\sigma}$, known as the \emph{weight function}, is chosen according to the aims of the simulation. 
Note that if $\eta_{\sigma}>0$ then state $\sigma$ is sampled more frequently than would be the case for the ensemble of interest; and if
$\eta_{\sigma}<0$ then $\sigma$ is sampled less frequently. The strength of this approach is that through judicious choice of the weight function, 
one can `control' the path the system traverses through phase space.

Of course, in a multicanonical simulation the states are no longer sampled with probabilities corresponding to the true ensemble in question -- 
which \emph{is} the case for canonical Monte Carlo. Accordingly the time average of some physical quantity $X$ throughout a long multicanonical 
Monte Carlo simulation is not equivalent to the equilibrium value of $X$ for the true ensemble, as it is in a canonical Monte Carlo
simulation. Nevertheless one can obtain the equilibrium value of $X$ from a multicanonical simulation by exploiting the fact that, since the 
weight function is known, then so is the degree of over- or under-sampling of each state. To elaborate, 
the equilibrium value of $X$ in a multicanonical Monte Carlo simulation is given by
\begin{equation}\label{equilX_MCMC}
\langle X\rangle\approx\frac{\displaystyle\sum_{t=1}^{\tau}e^{-\eta(t)}X(t)}{\displaystyle\sum_{t=1}^{\tau}e^{-\eta(t)}},
\end{equation}
where $X(t)$ denotes the quantity $X$ corresponding to the state sampled at timestep $t$, and $\tau$ denotes the total number of timesteps.

\subsection{Multicanonical Monte Carlo in LSMC}\label{sec:multicanonical_LSMC}
How does multicanonical Monte Carlo resolve the problem that lattice switch moves are too rarely accepted to be useful? Recall that lattice switches
are usually rejected because they result in a trial state with a much higher energy.
There are, however, a small number of states `close' to those realised at equilibrium for each phase from which a lattice switch yields a 
trial state $\sigma'$ which is of comparable energy to $\sigma$. From such states a lattice switch has a good chance of being accepted.
We refer to such states as \emph{gateway states}, since they provide the key to jumping between both phases. It is these states which we wish to over-sample,
and we set the weight function accordingly. The idea is that by over-sampling these states, lattice switches are accepted reasonably often, 
enabling both phases to be explored in a reasonable simulation time. This in turn allows us to determine $p_1$ and $p_2$,
and hence $\Delta \mathcal{F}$ via Eqn. \eqref{DeltaF_stat_mech}. Specifically, $p_1$ and $p_2$ are obtained via Eqn. \eqref{equilX_MCMC}:
\begin{equation}
p_{\alpha}=\langle\theta_{\alpha}\rangle \approx \frac{\displaystyle\sum_{t=1}^{\tau}e^{-\eta(t)}\theta_{\alpha}(t)}{\displaystyle\sum_{t=1}^{\tau}e^{-\eta(t)}},
\end{equation}
where $\theta_{\alpha}(t)$ takes the value 1 if the system is in phase $\alpha$ at timestep $t$ and 0 otherwise.

How should the weight function be engineered such that gateway states are over-sampled? 
Consider a state $\sigma$, and let $\sigma'$ denote the state which results from a lattice switch performed from $\sigma$. Let us define the 
state-dependent quantity
\begin{equation}\label{M_def}
M_{\sigma}=
\begin{cases}
(E_{\sigma}-E_{\sigma'}) & \text{if $\sigma$ belongs to phase 1} \\
-(E_{\sigma}-E_{\sigma'}) & \text{if $\sigma$ belongs to phase 2}.
\end{cases}
\end{equation}
This quantity provides a practical means for resolving gateway states, states corresponding to equilibrium (for the true ensemble under consideration)
for phase 1, and states corresponding to equilibrium for phase 2. Accordingly we refer to $M$ as the \emph{order parameter}. Consider first gateway states.
Above we illustrated that gateway states correspond to the condition $E_{\sigma}\approx E_{\sigma'}$. As can be seen from the above this corresponds to
states with $M_{\sigma}\approx 0$. By contrast, if $|M_{\sigma}|\gg 0$, then the two states have significantly different values of $E$. In this case, while
switching from the state with the higher value of $E$ to that with the lower value of $E$ is guaranteed, the converse is not: the two states are not 
concordant with switching \emph{to and from} both phases. $|E_{\sigma}|$ therefore provides a measure of how `un-gateway-like' state $\sigma$ is, with 
zero corresponding to `very gateway-like'. 
Consider now phase-1-equilibrium states. From such states we generally expect a lattice switch to be unsuccessful, and hence
$E_{\sigma'}\gg E_{\sigma}$. Therefore for such states $M_{\sigma}\ll 0$. Finally consider phase-2-equilibrium states. Similarly we generally expect 
a lattice switch from such states to be unsuccessful, and hence $E_{\sigma'}\gg E_{\sigma}$. However this time $M_{\sigma}\gg 0$. 
We therefore have three regimes: $M_{\sigma}\ll 0$ corresponds to phase-1-equilibrium states; $M_{\sigma}\approx 0$ corresponds to gateway states; and 
$M_{\sigma}\gg 0$ corresponds to phase-2-equilibrium states.

With this in mind, if we choose the weight function $\eta_{\sigma}$ to take the same value $\eta_{M}$ 
for all states with the same $M$ and also choose $\eta_{M}$ to be peaked at $M=0$ and to decay monotonically with 
$|M|$, then the weight function corresponds to a `force' which drives the system towards gateway states, allowing the system to transition between
the phase-1 and phase-2 regions of phase space, corresponding to $M\ll 0$ and $M\gg 0$ respectively, in a reasonable simulation time.
This is, of course, just a \emph{qualitative} description of a form for $\eta_{M}$ which is sufficient for our purposes. As one might expect, 
the quantitative details of the weight function $\eta_{M}$ strongly affect the efficiency of the path traversed though phase space with regards 
to calculating $\Delta \mathcal{F}$; a `bad' weight function might result in the system getting stuck in one phase, or an unimportant region of phase 
space, for a long time. Furthermore, it is not obvious \emph{a priori} what a suitable weight function for a given system should be. Hence one must 
\emph{generate} a weight function which leads to an efficient sampling of phase space. After this \emph{weight function generation simulation},
the resulting weight function can be used in a \emph{production simulation} to calculate $\Delta \mathcal{F}$ as described earlier.

However in practice one cannot treat $M$ as an unbounded continuous variable as above; one cannot define an arbitrary weight function in computer
memory via this scheme. Hence in practice one considers a finite range of $M$ which is divided into $N_{\text{macro}}$ bins, each 
corresponding to a distinct range of `$M$-space'. Each bin itself corresponds to a macrostate: the macrostate is the collection of states corresponding to the 
range of $M$-space covered by the bin. We will henceforth explicitly take this discretisation of $M$ into account, and let $\mathcal{M}$ denote the 
macrostate corresponding to the $\mathcal{M}$th bin, where $\mathcal{M}=1,2,\dotsc,N_{\text{macro}}$. Accordingly let 
$\eta_{\mathcal{M}}$ denote the weight function for macrostate $\mathcal{M}$.

\subsection{Weight function generation}\label{sec:weight_generation}
The weight function can be generated in many different ways, some of which are more efficient than others. We now list the methods implemented in
\emph{monteswitch}. All of these methods share the same notion of the `ideal' weight function $\eta_{M}^*$, which leads to all macrostates within
the considered order parameter range to be sampled with equal probability in the multicanonical Monte Carlo simulation.

\subsubsection{Visited states method}
The \emph{visited states method} (see Ref. \cite{Smith_1995} and references therein) is arguably the simplest method for generating the weight function. 
In the visited states method, the simulation consists of a number of `blocks', which themselves consist of a large number of Monte Carlo sweeps. 
Multicanonical sampling is used throughout, and the weight function is updated at the end of each block. The weight function is different -- closer to the 
ideal -- in each subsequent block, and the number of visits to all macrostates during each block is used to inform the weight function to 
be used in the next block. Eventually the weight function converges on the ideal: it provides a `flat' macrostate histogram; the weight function is 
such that all macrostates are sampled with equal probability.
Specifically, the following scheme is used to update the weight function at the end of each block:
\begin{equation}
\eta^{(n+1)}_{\mathcal{M}}=\eta^{(n)}_{\mathcal{M}}-\ln\Biggl\lbrace
\frac{\mathcal{C}^{(n)}_{\mathcal{M}}+1}{\sum_{\mathcal{M}'}\bigl(\mathcal{C}^{(n)}_{\mathcal{M}'}+1\bigr)}
\Biggr\rbrace
+k,
\end{equation}
where 
where $\mathcal{C}^{(n)}_{\mathcal{M}}$ denotes the number of states belonging to macrostate $\mathcal{M}$ visited during block $n$;
$\eta^{(n)}_{\mathcal{M}}$ denotes the weight function for block $n$; the summation over $\mathcal{M}'$ on the denominator of the fraction is
over all macrostates $1,2,\dotsc,N_{\text{macro}}$; and $k$ is an inconsequential arbitrary constant, which we choose such that the minimum value of 
$\eta^{(n+1)}_{\mathcal{M}}$ over all $\mathcal{M}$ is 0.

\subsubsection{Transition matrix method}\label{sec:transition_matrix}
A more sophisticated method than the visited states method, which is significantly more efficient, is the \emph{transition matrix method}
\cite{Smith_1995,Fitzgerald_1999}. This method exploits the fact that the ideal weight function $\eta_{\mathcal{M}}^*$ is related to the \emph{canonical}
probability $p_{\mathcal{M}}$ of the system being in macrostate $\mathcal{M}$ via the equation
\begin{equation}\label{ideal_wf}
\eta^*_{\mathcal{M}}=A-\ln p_{\mathcal{M}},
\end{equation}
where $A$ is an arbitrary constant. $p_{\mathcal{M}}$ in turn can be determined from the \emph{macrostate transition probability matrix} 
$\mathcal{T}_{\mathcal{MM}'}$, which describes the probability that the system, currently in macrostate $\mathcal{M}$, transitions to macrostate 
$\mathcal{M}'$ \emph{in the canonical ensemble}. In the transition matrix method we determine $\mathcal{T}_{\mathcal{MM}'}$, and then use this to obtain
$p_{\mathcal{M}}$, and finally the ideal weight function $\eta_{\mathcal{M}}^*$ via Eqn. \eqref{ideal_wf}.

$\mathcal{T}_{\mathcal{MM}'}$ is determined as follows. During the simulation we keep track of the number of transitions between all pairs of macrostates, 
which we store in a matrix $\mathcal{C}_{\mathcal{MM}'}$ -- where $\mathcal{C}_{\mathcal{MM}'}$ denotes the number of transitions from macrostate $\mathcal{M}$ to macrostate 
$\mathcal{M'}$. We then use $\mathcal{C}_{\mathcal{MM}'}$ to obtain an estimate for $\mathcal{T}_{\mathcal{MM'}}$ via the equation
\begin{equation}\label{T_estimate}
\mathcal{T}_{\mathcal{M}\mathcal{M}'}\approx \frac{\mathcal{C}_{\mathcal{M}\mathcal{M}'}+1}
{\displaystyle\sum_{\mathcal{M}^{\prime\prime}}\bigl(\mathcal{C}_{\mathcal{M}\mathcal{M}^{\prime\prime}}+1\bigr)}.
\end{equation}
However $\mathcal{C}_{\mathcal{MM'}}$ is not simply the number of \emph{observed} transitions from $\mathcal{M}$ to $\mathcal{M'}$ during the simulation, but 
rather the number of \emph{inferred} transitions. To elaborate, consider a trial state $\sigma'$ generated from a state $\sigma$ which, if accepted, would 
take the system from macrostate $\mathcal{M}$ to macrostate $\mathcal{M'}$, and let the \emph{canonical} probability of the move being accepted be 
$p_{\sigma\to\sigma'}$. Instead of performing the update $\mathcal{C}_{\mathcal{MM'}}\to \mathcal{C}_{\mathcal{MM'}}+1$ if the move is accepted and 
$\mathcal{C}_{\mathcal{MM'}}\to \mathcal{C}_{\mathcal{MM'}}$ if it is not -- which would result in $\mathcal{C}_{\mathcal{MM'}}$ being the number of observed transitions from 
$\mathcal{M}$ to $\mathcal{M'}$ -- we perform the update
\begin{equation}
\begin{split}
\mathcal{C}_{\mathcal{M}\mathcal{M}'}\to \mathcal{C}_{\mathcal{M}\mathcal{M}'}+p_{\sigma\to\sigma'} \\
\mathcal{C}_{\mathcal{M}\mathcal{M}}\to \mathcal{C}_{\mathcal{M}\mathcal{M}}+1-p_{\sigma\to\sigma'}
\end{split}
\end{equation}
regardless of whether it is accepted or not. Note that the canonical quantity $p_{\sigma\to\sigma'}$ is always used in the update procedure, which leads to
$\mathcal{C}_{\mathcal{MM'}}$ being the inferred number of canonical transitions between $\mathcal{M}$ and $\mathcal{M'}$. Because of this one can use any method 
for exploring $M$-space: canonical, multicanonical, or something else. We elaborate on this point in a moment.

Having determined  $\mathcal{T}_{\mathcal{MM}'}$, our task is to now calculate $p_{\mathcal{M}}$. It can be shown that the macrostates obey the following
detailed balance condition \cite{Smith_1995}:
\begin{equation}
\mathcal{T}_{\mathcal{M}'\mathcal{M}}p_{\mathcal{M}'}=\mathcal{T}_{\mathcal{MM}'}p_{\mathcal{M}}.
\end{equation}
Setting $\mathcal{M}'=\mathcal{M}+1$ and rearranging the above gives
\begin{equation}\label{shooting}
p_{(\mathcal{M}+1)}=\frac{\mathcal{T}_{\mathcal{M}(\mathcal{M}+1)}}{\mathcal{T}_{(\mathcal{M}+1)\mathcal{M}}}p_{\mathcal{M}}.
\end{equation}
Using this equation, $p_{\mathcal{M}}$ can be obtained from the matrix $\mathcal{T}_{\mathcal{M}\mathcal{M}'}$ via the following procedure.
Firstly, one chooses some arbitrary value for $p_1$. 
\footnote{In this section $p_1$ and $p_2$ denote the probability of the system being in macrostates 1 and 2, \emph{not} the probabilities of
the system being in phases 1 and 2.}
With this $p_2$ can be obtained from the above equation ($\mathcal{M}=1$ in Eqn. \eqref{shooting}). 
This in turn can be used to obtain $p_3$ ($\mathcal{M}=2$ in Eqn. \eqref{shooting}), which in turn can be used to obtain $p_4$, etc., until
$p_{N_{\text{macro}}}$ is obtained. Finally, one normalises the resulting function $p_{\mathcal{M}}$ such that
\begin{equation}
\sum_{\mathcal{M}=1}^{N_{\text{macro}}}p_{\mathcal{M}}=1,
\end{equation}
as is required. The final step is to use $p_{\mathcal{M}}$ to obtain an estimate for the ideal weight function. This is done simply by substituting 
$p_{\mathcal{M}}$ into Eqn. \eqref{ideal_wf}.

\subsubsection{Methods for exploring $M$-space}\label{sec:exploring_M_space}
As alluded to above, since the updates to $\mathcal{C}_{\mathcal{M}\mathcal{M}'}$ always use the canonical probabilities of transitioning between states, 
with the transition matrix method one can \emph{choose} how $M$-space is explored. \emph{monteswitch} supports a number of ways of doing this.

The first method is to use multicanonical sampling to explore $M$-space with a continuously evolving weight function, where the weight function
at a given time is the current estimate for the ideal weight function derived from the current $\mathcal{C}_{\mathcal{M}\mathcal{M}'}$ as described above.
This is the `natural' way of applying the transition state method.

The second method is to use what we refer to as \emph{artificial dynamics} to force the system to explore all macrostates in a reasonable amount of time. 
In this method, the system is first locked into a macrostate for a certain period of time. After that period of time has elapsed, the `barriers' preventing
the system from moving into an adjacent macrostate is moved such that the system is free to transition into an adjacent macrostate. Once this occurs,
the system is locked into this new macrostate, and the procedure starts again. There is of course the question of which adjacent macrostate to `open'
to the system. Assuming we are not in macrostate $\mathcal{M}=1$ or $N_{\text{macro}}$, then there are two options: $(\mathcal{M}+1)$ and $(\mathcal{M}-1)$.
In \emph{monteswitch} one can specify whether to select the new macrostate at random \cite{thesis:Jackson}, 
or whether to sweep through the macrostates systematically, e.g.,
to explore macrostates $3,4,5,\dotsc,(N_{\text{macro}}-1),N_{\text{macro}},(N_{\text{macro}}-1),\dotsc,3,2,1,2,3,\dotsc$. This method is faster than the 
`natural' method just described because one does not have to wait for the weight function to evolve such that it pushes the system to explore macrostates 
which are unlikely to be visited in the canonical ensemble.

\section{Package structure}\label{sec:preliminaries}
The \emph{monteswitch} package consists of a number of programs, as well as a user manual, a suite of test cases, a worked example, and a suite
of Fortran modules corresponding to different interatomic potentials. The programs are:
\begin{itemize}
\item\texttt{monteswitch}
\item\texttt{monteswitch\_mpi}
\item\texttt{monteswitch\_post}
\item\texttt{lattices\_in\_hcp\_fcc}
\item\texttt{lattices\_in\_bcc\_fcc}
\item\texttt{lattices\_in\_bcc\_hcp}
\end{itemize}
\texttt{monteswitch} and \texttt{monteswitch\_mpi} are the key programs of the package: they perform Monte Carlo simulations. 
By contrast \texttt{monteswitch\_post}, \texttt{lattices\_in\_hcp\_fcc}, \texttt{lattices\_in\_bcc\_fcc} and 
\texttt{lattices\_in\_bcc\_hcp} are utility programs: \texttt{monteswitch\_post} is for post-processing one of the output files created 
by the main programs; and \texttt{lattices\_in\_hcp\_fcc}, \texttt{lattices\_in\_bcc\_fcc} and \texttt{lattices\_in\_bcc\_hcp} are for generating 
one of the input files for the main programs. We elaborate upon these programs in later sections.

\section{Interatomic potentials}\label{sec:potentials}
The file \texttt{interactions.f95} in the main directory of the package contains the Fortran module, named \texttt{interactions\_mod}, which
determines the interatomic potential to be utilised by the \emph{monteswitch} programs \texttt{monteswitch}, \texttt{monteswitch\_mpi} and
\texttt{monteswitch\_post} after the package is compiled. 
By default \texttt{interactions.f95} corresponds to the embedded atom model (EAM) \cite{Daw_1984}; 
to implement a specific interatomic potential one must copy 
the corresponding \texttt{interactions.f95} file to \texttt{interactions.f95} in the main directory of the package, and then compile the package.

\subsection{Structure of \texttt{interactions\_mod}}
The module \texttt{interactions\_mod} contains the following procedures which interface with the main \emph{monteswitch} programs
\texttt{monteswitch} and \texttt{monteswitch\_mpi}:
\begin{itemize}
\item
\texttt{initialise\_interactions}, which initialises the variables within the module, possibly 
by reading variables from one or more input files, for `new' simulations; 
\item
\texttt{export\_interactions}, which exports the module variables to a file for the purposes
of checkpointing the simulation; 
\item
\texttt{import\_interactions}, which imports the module variables from the aforementioned file to resume a checkpointed
simulation; 
\item
\texttt{after\_accepted\_part\_interactions}, \texttt{after\_accepted\_vol\_interactions} and 
\texttt{after\_accepted\_lattice\_interactions}, which perform any housekeeping tasks required by the module (e.g., updating neighbour lists)
after, respectively, a particle, volume and lattice switch move has been accepted; 
\item
\texttt{after\_all\_interactions}, which performs any housekeeping tasks required by the module after all moves, including
rejected moves; 
\item
\texttt{calc\_energy\_scratch}, which calculates the energy of the system `from scratch' for a specified state;
\item
\texttt{calc\_energy\_part\_move}, which calculates the energy of the system given that one particle has moved.
\end{itemize}
Users wishing to write their own versions of \texttt{interactions.f95} to implement their own interatomic potentials must write
their own versions of each of the above procedures. To assist with this, two templates for \texttt{interactions.f95} are provided
with \emph{monteswitch}. These can be found in the directory \texttt{Interactions} within the package. 
The file \texttt{interactions\_TEMPLATE\_minimal.f95} contains a `bare' version of \texttt{interactions.f95}, i.e., it leaves 
the above procedures empty to be filled in by the user. The file \texttt{interactions\_TEMPLATE\_pair.f95}
allows quick generation of \texttt{interactions.f95} files for pair potentials, something which we elaborate upon below. Both templates contain comments
which provide guidance to the user.

Note that there is freedom in how the module variables are initialised for a new simulation -- via the procedure \texttt{initialise\_interactions}.
Normally \texttt{initialise\_interactions} would read the variables -- which parametrise the potential under consideration -- from one or more 
input files. We emphasise that these input files, and their formats, depend on the specific version of \texttt{interactions.f95} which 
\emph{monteswitch} is used in conjunction with.

\subsection{Potentials included with \emph{monteswitch}}
The \texttt{Interactions} directory within the package contains a number of versions of \texttt{interactions.f95}, which correspond
to various interatomic potentials. The most important of these are as follows.

\subsubsection{Embedded atom model}\label{sec:EAM}
The file \texttt{interactions\_EAM.f95} implements the embedded atom model (EAM) \cite{Daw_1984} for metals (but not alloys). 
Here the energy of the system is given by
\begin{equation}
E = \frac{1}{2}\sum_{i,j\neq i}\phi(r_{ij}) + \sum_iF(\rho_i),
\end{equation}
\begin{equation}
\rho_i=\sum_{j\neq i}\rho(r_{ij}),
\end{equation}
where $r_{ij}$ is the separation between particles $i$ and $j$, and $\phi$, $F$ and $\rho$ are functions which must be specified
and constitute the parametrisation of the EAM potential. If \texttt{interactions\_EAM.f95} is used then one input file, named
\texttt{interactions\_in}, is required by the programs \texttt{monteswitch} and \texttt{monteswitch\_mpi} to input the potential for new simulations.
This file must be a description of the EAM potential to be used in DYNAMO/LAMMPS `setfl' format \cite{website:LAMMPS_EAM}.

\subsubsection{Soft pair potentials}\label{section:pair_potentials}
Table \ref{table:pair_potentials} gives a list of soft pair potentials included with \emph{monteswitch}, along with the name 
of the corresponding \texttt{interactions.f95} file in \texttt{Interactions}. All of these potentials are implemented in the same way.
Firstly, the pair potential $\phi(r)$ is assumed to be 0 for inter-particle separations $r$ greater than some cut-off distance $r_{\text{c}}$. In other
words the potential is \emph{truncated} at $r_{\text{c}}$. Secondly, only pairs of particles within a distance $r_{\text{list}}$ of each other at 
the start of the simulation interact with each other throughout the entire simulation.
To clarify the difference between $r_{\text{c}}$ and $r_{\text{list}}$: the former is the distance at which the potential is truncated, while
the latter determines which particles are in each others' \emph{neighbour list}, which remains constant throughout the simulation. 

As for the EAM potential just described, for the soft potentials an input file \texttt{interactions\_in}, is required by the programs 
\texttt{monteswitch} and \texttt{monteswitch\_mpi} to input the potential for new simulations.
The format of this file is as follows: each variable which parametrises the potential corresponds to a specific line in \texttt{interactions\_in}, 
and each line must contain a string (we recommend the name of the variable followed immediately by an `=' character with no spaces), followed 
by whitespace, followed by the value of the variable. The final column of Table \ref{table:pair_potentials} gives the order of variables as 
they should appear, one per line, in \texttt{interactions\_in} for each potential. Note that in all cases $r_{\text{c}}$ and $r_{\text{list}}$ are
included as input variables. Furthermore there is a variable $N_{\text{list}}$, which determines the size of the array used in the program to store
the neighbour lists. Details regarding this variable can be found in the user manual.
To illustrate the above, here is an example \texttt{interactions\_in} file for the Lennard-Jones potential (\texttt{interactions\_LJ.f95}):
\begin{verbatim}
lj_epsilon= 1.0
lj_sigma= 1.0
lj_cutoff= 1.5
list_cutoff= 1.000000001
list_size= 14
\end{verbatim}

\begin{landscape}
\begin{center}\label{table:pair_potentials}
\begin{longtable}{ l l l p{3cm} }
\caption{Soft interatomic pair potentials included with \emph{monteswitch}}
\\
Potential & File name & Expression for potential & Order of variables in \texttt{interactions\_in}  \\

\hline

12-10 & \texttt{interactions\_12-10.f95} & $A/r^{12}-B/r^{10}$ 
& $A$, $B$, $r_{\text{c}}$, $r_{\text{list}}$, $N_{\text{list}}$ \\

12-6  &  \texttt{interactions\_12-6.f95} & $A/r^{12}-B/r^{6}$ 
& $A$, $B$, $r_{\text{c}}$, $r_{\text{list}}$, $N_{\text{list}}$ \\

Buckingham & \texttt{interactions\_Buckingham.f95} & $A\exp(-r/\rho)-C/r^6$ 
&  $A$, $\rho$, $C$, $r_{\text{c}}$, $r_{\text{list}}$, $N_{\text{list}}$ \\

Gaussian & \texttt{interactions\_Gaussian.f95} & $-A\exp(-Br^2)$ 
& $A$, $B$, $r_{\text{c}}$, $r_{\text{list}}$, $N_{\text{list}}$ \\

Lennard-Jones & \texttt{interactions\_LJ.f95} & $4\epsilon\bigl[(\sigma/r)^{12}-(\sigma/r)^6\bigr]$
& $\epsilon$, $\sigma$, $r_{\text{c}}$, $r_{\text{list}}$, $N_{\text{list}}$ \\

9-6 Lennard-Jones & \texttt{interactions\_LJ\_9-6.f95} & $4\epsilon\bigl[(\sigma/r)^9-(\sigma/r)^6\bigr]$
& $\epsilon$, $\sigma$, $r_{\text{c}}$, $r_{\text{list}}$, $N_{\text{list}}$ \\

Morse & \texttt{interactions\_Morse.f95} & $E_0\bigg\lbrace \Bigl[1-\exp\bigl(-k(r-r_0)\bigr)\Bigr]^2 -1 \biggr\rbrace$ 
& $E_0$, $k$, $r_0$, $r_{\text{c}}$, $r_{\text{list}}$, $N_{\text{list}}$ \\

$n$-$m$ & \texttt{interactions\_n-m.f95} & $\displaystyle\bigl[E_0/(n-m)\bigr]\bigl[m(r_0/r)^n-n(r_0/r)^m\bigr]$
& $n$, $m$, $r_0$, $r_{\text{c}}$, $r_{\text{list}}$, $N_{\text{list}}$ \\

Yukawa & \texttt{interactions\_Yukawa.f95} & $A\exp(-kr)/r$ 
& $A$, $k$, $r_{\text{c}}$, $r_{\text{list}}$, $N_{\text{list}}$ \\

\end{longtable}
\end{center}
\end{landscape}

\subsubsection{User-defined pair potentials}\label{sec:user_defined}
As mentioned above, the file \texttt{interactions\_TEMPLATE\_pair.f95} is a template which can be used to easily create \texttt{interactions.f95} 
files for user-defined pair potentials. Instructions are provided in the file regarding how the file should be modified to realise the user's 
potential of interest. In fact this template was used to create the files for almost all of the pair potentials described above.

\subsubsection{Hard/penetrable spheres}\label{sec:penetrable_spheres}
The file \texttt{interactions\_HS\_multi.f95} implements the penetrable (including hard) spheres model, where the sphere diameter is allowed to 
vary with particle species. Here, the pair potential between two particles belonging to species $s$ and $t$ is given by
\begin{equation}
\phi_{st}(r)=
\begin{cases} 
\epsilon & \text{if }r<\frac{1}{2}(\sigma_s+\sigma_t) \\
0 & \text{otherwise},
\end{cases}
\end{equation}
where $\sigma_s$ denotes the diameter of spheres belonging to species $s$. Again, the input file for the new simulations in the programs
\texttt{monteswitch} and \texttt{monteswitch\_mpi} is named \texttt{interactions\_in}, and its format is similar to the \texttt{interactions\_in}
files for the soft pair potentials described above. In this case however the order of the variables is
$\epsilon$, $N_{\text{species}}$, $\mathbf{\sigma}$, $r_{\text{list}}$, and $N_{\text{list}}$, where $N_{\text{species}}$ is the number of species to
consider, and $\mathbf{\sigma}$ is an $N_{\text{species}}$-dimensional vector (to be specified on a single line \texttt{interactions\_in}) 
containing the diameters for each species $1,2,\dotsc,N_{\text{species}}$. Note that hard spheres correspond to the limit $\beta\epsilon\to\infty$, 
and hence hard spheres can be implemented by setting $\epsilon$ to a high value.

\section{Monte Carlo simulation programs}\label{sec:simulation_programs}
The key programs in the package are \texttt{monteswitch} and \texttt{monteswitch\_mpi}.
The main purpose of these programs is to perform LSMC simulations, though they can also
be used to perform `conventional' Monte Carlo simulations. \texttt{monteswitch\_mpi} is the MPI-parallelised analogue of \texttt{monteswitch}. 
While in \texttt{monteswitch} one simulation is performed, in \texttt{monteswitch\_mpi} multiple simulations are 
performed in parallel. These simulations are identical except for the seed used for the random number generator. Hence \texttt{monteswitch\_mpi}
is simply a convenient means to exploit parallelisation in order to obtain results quickly.
In terms of usage and simulation methodology \texttt{monteswitch} and \texttt{monteswitch\_mpi} are almost identical. For this reason we 
focus on \texttt{monteswitch} below, where it should be assumed that what is said for \texttt{monteswitch\_mpi} also applies for
\texttt{monteswitch\_mpi} unless otherwise stated.

\subsection{Overview of functionality}
\texttt{monteswitch} can treat phases which can be represented by orthorhombic supercells in the NVT and NPT ensembles, where in the
NPT ensemble both isotropic volume moves and volume moves which allow the shape of the system to alter are supported. Multicomponent
systems are allowed, however \texttt{monteswitch} cannot treat `molecular' systems in which the particles have orientational degrees of
freedom: rotational Monte Carlo moves are not implemented in \texttt{monteswitch}.

\texttt{monteswitch} supports both canonical and multicanonical sampling, where the multicanonical
sampling is performed over the LSMC order parameter described in Section \ref{sec:multicanonical_LSMC}.
Regarding weight function generation, \texttt{monteswitch} supports all of the methods described in Section \ref{sec:weight_generation}. 
Note that the option to perform canonical sampling enables conventional Monte Carlo simulations to be performed. 
Note also that, like all other LSMC implementations the authors are aware of, \texttt{monteswitch} explicitly keeps track of two states of 
the system during a simulation, one corresponding to phase 1 and one corresponding to phase 2. At any point during the simulation only one 
of these is the `actual' state of the system $\sigma$, while the other is the state $\sigma'$ which would result if a lattice switch were
performed from $\sigma$. For instance if the system were in phase 1, then $\sigma$ would belong to phase 1 and $\sigma'$ would belong to 
phase 2. During a simulation the actual state $\sigma$ is evolved via particle moves, and additionally volume moves for the case of the NPT ensemble, 
in the conventional manner. However the `other' state $\sigma'$ is evolved contemporaneously with $\sigma$ so that $\sigma'$ is always what 
would result if a lattice switch were performed from $\sigma$. It is necessary to continuously track $\sigma'$ in this manner because the LSMC order 
parameter for $\sigma$ depends on the energy, and possibly also the volume, of $\sigma'$ (see Section \ref{sec:multicanonical_LSMC} for details). 
Regarding lattice switch moves, if such a move is accepted, then the actual state is simply relabelled from $\sigma$ to $\sigma'$, while $\sigma$ is 
relabelled as the `other' state.

\texttt{monteswitch} also supports on-the-fly evaluation of physical quantities and their uncertainties during the simulation via block 
averaging (see, e.g., Ref. \cite{book:Frenkel}). It also supports the ability to check whether or not 
the system has `melted', i.e., whether or not one or more of the particles have strayed `too far' from their lattice sites. In a similar vein, 
it is possible in \texttt{monteswitch} to perform simulations in the centre-of-mass reference frame, which provides a means of suppressing spurious
melting due to `drift' in the centre-of-mass of the system during the simulation.

The random number generator utilised by \texttt{monteswitch} is the Mersenne Twister (MT19937) \cite{Matsumoto_1998}.

\subsection{Command-line argument usage}
The command-line arguments passed to \texttt{monteswitch} determine the nature of the invoked simulation. Usage of \texttt{monteswitch} is
as follows: 
\begin{verbatim}
monteswitch [-seed <seed>] -new [-wf]
monteswitch [-seed <seed>] (-resume|-reset)
\end{verbatim}
The function of each of these arguments is described below.

\subsection{Seeding the random number generator}
The command-line argument \texttt{-seed} allows the user to specify the seed for the forthcoming simulation explicitly.
If the argument \texttt{-seed} is absent then a seed is generated using the system clock.

\subsection{Running a new simulation}
The command-line argument \texttt{-new} invokes a new simulation `from scratch'. In this case
the simulation is initialised using information contained in input files located in the current directory. The input files required
by \texttt{monteswitch} will depend on the specific version of \texttt{interactions.f95} utilised when compiling \emph{monteswitch} 
(see Section \ref{sec:potentials}). 
However it is only the input files which contain information pertaining to the interatomic potential which are version-dependent;
the remaining information used to initialise a simulation are read from input files which are universal to all versions of 
\texttt{monteswitch}, namely \texttt{lattices\_in}, \texttt{params\_in} and \texttt{wf\_in}. 
The first two of these are compulsory: they are read by all new simulations. By contrast \texttt{wf\_in} is optional, only being read if 
the command-line argument \texttt{-wf} is present.

\subsubsection{Input file: \texttt{lattices\_in}}
We now describe these files, beginning with \texttt{lattices\_in}. 
This file contains specifications for two states (i.e., supercell dimensions and particle positions), one for
each phase. These two states serve two purposes. Firstly, they act as prospective initial state for the simulation: if the system is to be 
initialised in phase 1, then the phase-1 state will be used as the initial state, and similarly if the system is to be initialised in phase 2. 
Secondly, they determine the nature of the lattice switch. The fractional position of particle $i$ for phase $\alpha$ specified in \texttt{lattices\_in}
is used as the lattice site for $i$ in phase $\alpha$ during lattice switches. Furthermore the supercell dimensions specified for each phase in
\texttt{lattices\_in} determine how the supercell is transformed during lattice switches.
Specifically, if $L^{(\alpha)}_x$, $L^{(\alpha)}_y$ and $L^{(\alpha)}_z$ denote the supercell dimensions specified for phase $\alpha$ in
\texttt{lattices\_in}, then the $x$-, $y$- and $z$-dimensions of the supercell are multiplied by factors $L^{(2)}_x/L^{(1)}_x$, 
$L^{(2)}_y/L^{(1)}_y$ and $L^{(2)}_z/L^{(1)}_z$ respectively during a lattice switch from phase 1 to phase 2, and by $L^{(1)}_x/L^{(2)}_x$, 
$L^{(1)}_y/L^{(2)}_y$ and $L^{(1)}_z/L^{(2)}_z$ during a lattice switch from phase 2 to phase 1.
Below is a pedagogical example of a \texttt{lattices\_in} file, which corresponds to phase 1 being an 8-particle bcc supercell and phase 2 being 
an 8-particle hcp supercell, where the phase-1 state consists entirely of particles belonging to species `1'and the phase-2 state consists of a 
mixture of species `1' and `2'. Note that in this case the species of some of the particles is transformed during lattice switches.
\begin{verbatim}
 bcc-hcp, rho = 0.5, nx,ny,nz = 1, 1, 2   # Comment line
   8                                      # Number of particles
 2.2449241                                # x-dimension for phase 1
 1.5874012                                # y-dimension for phase 1
 4.4898482                                # z-dimension for phase 1
   0.0000000   0.0000000   0.0000000   1  # Coords and species for phase 1
   0.5000000   0.5000000   0.0000000   1
   0.5000000   0.0000000   0.2500000   1
   0.0000000   0.5000000   0.2500000   1
   0.0000000   0.0000000   0.5000000   1
   0.5000000   0.5000000   0.5000000   1
   0.5000000   0.0000000   0.7500000   1
   0.0000000   0.5000000   0.7500000   1
 2.4494897                                # x-dimension for phase 2
 1.4142136                                # y-dimension for phase 2
 4.6188021                                # z-dimension for phase 2
   0.0000000   0.0000000   0.0000000   1  # Coords and species for phase 2
   0.5000000   0.5000000   0.0000000   2
   0.3333333   0.0000000   0.2500000   1
   0.8333333   0.5000000   0.2500000   2
   0.0000000   0.0000000   0.5000000   1
   0.5000000   0.5000000   0.5000000   2
   0.3333333   0.0000000   0.7500000   1
   0.8333333   0.5000000   0.7500000   2
\end{verbatim}

\subsubsection{Input file: \texttt{params\_in}}
The second compulsory input file for a new simulation is \texttt{params\_in}. This file contains the variables which determine the nature 
of the simulation. Each variable corresponds to a specific single line in the file, and each line must consist of an arbitrary string (we 
recommend the name of the variable followed immediately by an `=' character with no spaces), followed by whitespace, followed by the 
value of the variable. To illustrate this, below is an excerpt from a \texttt{params\_in} file:
\begin{verbatim}
init_lattice=                            1
M_grid_size=                             100
M_grid_min=                              -82.0
M_grid_max=                              48.0
enable_multicanonical=                   T
beta=                                    9.403
P=                                       0.0
enable_lattice_moves=                    T
enable_part_moves=                       T
enable_vol_moves=                        T
part_select=                             "rand"
part_step=                               0.3
enable_COM_frame=                        T
vol_dynamics=                            "UVM"
vol_freq=                                1
vol_step=                                0.03
stop_sweeps=                             160000
equil_sweeps=                            0
enable_melt_checks=                      T
melt_sweeps=                             100
melt_threshold=                          3.0
melt_option=                             "zero_current"
\end{verbatim}
A full list of the variables which must appear in a \texttt{params\_in}, as well as a description of their function, is provided in the
user manual included with the package. The key variables are listed in Table \ref{table:params_in_variables}.
Numerous examples of \texttt{params\_in} files are included with \emph{monteswitch} which serve as templates for users.

\begin{landscape}
\begin{center}\label{table:params_in_variables}
\begin{longtable}{l l p{13cm}}
\caption{Important control variables for \texttt{monteswitch} to be specified in the \texttt{params\_in} input file.}
\\

Variable & Type & Description \\
\hline
\textbf{init\_lattice}  &  \texttt{INTEGER}  & Starting phase for the simulation (1 or 2). \\
\textbf{M\_grid\_size}  &  \texttt{INTEGER}  & Number of macrostates to divide the considered order parameter range (\textbf{M\_grid\_min} to 
\textbf{M\_grid\_max}) into. \\
\textbf{M\_grid\_min}  &  \texttt{REAL}  & Minimum of considered order parameter range. \\
\textbf{M\_grid\_max}  &  \texttt{REAL}  & Maximum of considered order parameter range. \\
\textbf{enable\_multicanonical}  &  \texttt{LOGICAL}  & \texttt{T} enables multicanonical sampling using the current weight function; \texttt{F} 
enables canonical sampling. \\
\textbf{beta}  &  \texttt{REAL}  & Inverse temperature: $\beta=1/(k_BT)$. \\
\textbf{P}  &  \texttt{REAL}  & Pressure (only relevant in NPT ensemble simulations).  \\
\textbf{enable\_lattice\_moves}  &  \texttt{LOGICAL}  & \texttt{T} enables lattice switch moves (performed after every particle and volume move). \\
\textbf{enable\_part\_moves}  &  \texttt{LOGICAL}  & \texttt{T} enables particle moves. \\
\textbf{enable\_vol\_moves}  &  \texttt{LOGICAL}  & \texttt{T} enables volume moves and selects the NPT ensemble; \texttt{F} selects the NVT ensemble. 
A volume move will be attempted on average \textbf{vol\_freq} times per sweep. \\
\textbf{part\_select}  &  \texttt{CHARACTER(30)}  & Flag determining how the next particle to move is selected: \texttt{"cycle"} selects particles 
sequentially, \texttt{"rand"} selects particles at random.  \\
\textbf{part\_step}  &  \texttt{REAL}  & Particle move maximum size; particles are moved according to a random walk, with a maximum move size of 
\textbf{part\_step} in any Cartesian direction.  \\
\textbf{enable\_COM\_frame}  &  \texttt{LOGICAL}  & \texttt{T} performs the simulation in the centre-of-mass reference frame; \texttt{F} uses 
the lab frame. Using the centre-of-mass frame prevents `drift' in the centre-of-mass, which is convenient because it keeps particles close to
their lattice sites. \\
\textbf{vol\_dynamics}  &  \texttt{CHARACTER(30)}  & Flag determining which type of volume moves are performed: \texttt{"FVM"} (fixed volume move) 
keeps the supercell shape unchanged during a volume move; \texttt{"UVM"} (unconstrained volume move) allows the x-, y- and z-dimensions to move 
independently. \\
\textbf{vol\_freq}  &  \texttt{INTEGER}  & Number of volume moves performed per sweep on average if volume moves are enabled.
We recommend that this be set to 1.\\
\textbf{vol\_step}  &  \texttt{REAL}  & Volume move maximum step size; the volume is moved according to a random walk in `$\ln(V)$-space', with a 
maximum move size of \textbf{vol\_step}.  \\
\textbf{stop\_sweeps}  &  \texttt{INTEGER}  & Total number of Monte Carlo sweeps to perform in the simulation. \\
\textbf{equil\_sweeps}  &  \texttt{INTEGER}  & Number of sweeps to disregard before the system is considered to be equilibrated; statistics are not 
gathered during these sweeps for block averaging (see below). \\
\textbf{enable\_melt\_checks}  &  \texttt{LOGICAL}  & \texttt{T} enables periodic checks of whether the system has `melted', i.e., if one or 
more of the particles has moved more than a distance of \textbf{melt\_threshold} from its lattice site in any Cartesian direction then the system is 
considered to have `melted'. \\
\textbf{melt\_sweeps}  &  \texttt{INTEGER}  & Period (sweeps) to check for melting.  \\
\textbf{melt\_threshold}  &  \texttt{REAL}  & See \textbf{enable\_melt\_checks}.  \\
\textbf{melt\_option}  &  \texttt{CHARACTER(30)}  & Flag determining what the simulation does if the system has `melted': \texttt{"zero\_1"} and 
\texttt{"zero\_2"} move the system to the zero-displacement states in phases 1 and 2, respectively; \texttt{"zero\_current"} does the same but for 
the current phase; \texttt{"stop"} stops the simulation. For \texttt{"zero\_1"}, \texttt{"zero\_2"}, \texttt{"zero\_current"} the 
system is allowed to re-equilibrate before statistics are gathered for bock averaging. Also, the current block is disregarded for the purposes of 
block averaging. \\
\textbf{enable\_divergence\_checks}  &  \texttt{LOGICAL}  & \texttt{T} enables periodic checks of whether the energy of the system is correct,
given that during particle moves the energy of the system is \emph{amended}, as opposed to being calculated from scratch every move. If the
energy of the system differs from its `true' energy by an amount \textbf{divergence\_tol} then the simulation is stopped. \\
\textbf{divergence\_sweeps}  &  \texttt{INTEGER}  & Period (sweeps) to perform energy checks as just mentioned. \\
\textbf{divergence\_tol}  &  \texttt{REAL}  & See \textbf{enable\_divergence\_checks}.  \\
\textbf{output\_file\_period}  &  \texttt{INTEGER}  & Period (sweeps) at which information about the simulation is output to the file \texttt{data}. \\
\textbf{output\_stdout\_period}  &  \texttt{INTEGER}  & Period (sweeps) at which information about the simulation is output to stdout. \\
\textbf{checkpoint\_period}  &  \texttt{INTEGER}  & Period (sweeps) at which the simulation is checkpointed, i.e., how often all 
simulation variables are output to the file \texttt{state}. \\
\textbf{update\_eta}  &  \texttt{LOGICAL}  & \texttt{T} results in the weight function being periodically updated every 
\textbf{update\_eta\_sweeps} sweeps, according to the method specified in \textbf{update\_eta\_method}; 
\texttt{F} results in the weight function not being updated -- it remains frozen at its current state.  \\
\textbf{update\_eta\_sweeps}  &  \texttt{INTEGER}  & Period (sweeps) at which the weight function is updated. \\
\textbf{update\_trans}  &  \texttt{LOGICAL}  & \texttt{T} results in the transition matrix being updated; 
\texttt{F} results in it not being updated.  \\
\textbf{update\_eta\_method}  &  \texttt{CHARACTER(30)}  & Method used to update the weight function: \texttt{"VS"} uses the visited 
states method; \texttt{"shooting"} uses the transition matrix method.  \\
\textbf{enable\_barriers}  &  \texttt{LOGICAL}  & \texttt{T} enables artificial dynamics; for \texttt{F} the system is free to explore 
any macrostate, but is constrained to reside within the considered order parameter range (\textbf{M\_grid\_min} to \textbf{M\_grid\_max}). \\
\textbf{barrier\_dynamics}  &  \texttt{CHARACTER(30)}  & Flag determining how the macrostate barriers will evolve during artificial dynamics. 
All methods lock the system into a single macrostate for \textbf{lock\_moves} moves, before unlocking an adjacent macrostate. Once 
the system has moved into the adjacent macrostate, the system is then locked into that macrostate, and the procedure starts again. 
\texttt{"random"} evolves the macrostate the system is locked into via a random walk: the next macrostate is decided with equal probability 
to be that above or that below the current macrostate. \texttt{"pong\_up"} moves to increasingly higher macrostates until the upper limit of 
the supported order parameter range is encountered, at which point it reverses direction and proceeds to increasingly lower macrostates until 
it reaches the lower limit of the order parameter range, at which point it reverses direction, etc. \texttt{"pong\_down"}
instead moves initially to increasingly lower macrostates.  \\
\textbf{lock\_moves}  &  \texttt{INTEGER}  & The number of moves to lock the system into one macrostate for if artificial dynamics is used. \\
\textbf{calc\_equil\_properties}  &  \texttt{LOGICAL}  & \texttt{T} enables internal calculation of various physical quantities, and associated
uncertainties, via block averaging. \\
\textbf{block\_sweeps}  &  \texttt{INTEGER}  & The number of sweeps which comprise a `block' which will be used to evaluate physical quantities and
their associated uncertainties via block averaging. \\
\end{longtable}
\end{center}
\end{landscape}

\subsubsection{Input file: \texttt{wf\_in}}
The command-line argument \texttt{-wf} allows one to specify the initial weight function to be used in the simulation: if \texttt{-wf} is present, then
the initial weight function is read from the file \texttt{wf\_in}. If \texttt{-wf} is absent then \texttt{wf\_in} is not read, and the weight function 
is initialised to 0 for all macrostates. \texttt{wf\_in} must contain \textbf{M\_grid\_size} lines
(where \textbf{M\_grid\_size} is specified in \texttt{params\_in}), each containing two tokens (extra lines and tokens are ignored),
which both should be of type \texttt{REAL}. The first token on each line is ignored, while the second tokens are treated as the
weight function: the value of the weight function for macrostate $i$ is initialised to the value of the second token on line $i$ in
\texttt{wf\_in}. Note that the format of \texttt{wf\_in} is analogous to that output by the program \texttt{monteswitch\_post} in conjunction with the
\texttt{-extract\_wf} argument -- see Section \ref{sec:monteswitch_post}.

\subsection{Simulation output}
During a simulation information is periodically output to a file \texttt{data} and (optionally) stdout. Exactly what information is output is
controlled by flags in the input file \texttt{params\_in}. \texttt{data} can be used to deduce how the system evolves with time during the simulation.
The format of this file is transparent: each line contains a simulation variable (e.g., energy, volume), followed by the sweep number, followed 
by the value of the variable. To illustrate this, below is an excerpt from a \texttt{data} file:
\begin{verbatim}
 Lx:          250   20.506880155055160        22.375541637220159     
 Ly:          250   21.983483940969180        19.585057663268277     
 Lz:          250   20.142474063147095        20.720990682970093     
 V:          250   9080.4825242547813     
 lattice:          250           2
 E:          250  -2420.3817246101821     
 M:          250   12.808541584236991     
 eta:          250   9.9865126253137664     
 barrier_macro_low:          250          84
 Lx:          500   20.470022712854778        22.335325610873863     
 Ly:          500   21.965084799027686        19.568665891297709     
 Lz:          500   20.294516764376276        20.877400237511747     
 V:          500   9124.9380258152232     
 lattice:          500           2
 E:          500  -2409.3469532112986     
 M:          500   1.1974997526594968     
 eta:          500   54.025947498280964     
 barrier_macro_low:          500          84
 Lx:          750   20.472567199702674        22.338101960615028     
 Ly:          750   21.653559219100391        19.291128151472556     
 Lz:          750   20.503721368631055        21.092613455213190     
 V:          750   9089.3805950278784     
 lattice:          750           1
\end{verbatim}

In addition to \texttt{data}, a file \texttt{state} is also created by the program periodically throughout a simulation. This file contains a
snapshot of all the simulation variables, and can be used for checkpointing (discussed in a moment), or to extract the `results' of the simulation, 
e.g., equilibrium quantities, the current weight function, the number of accepted vs. rejected Monte Carlo moves of a certain type. The format of this
file is transparent. Each line pertains to a simulation variable, and the line itself contains the name of the simulation variable followed by the
variables value. To illustrate this, below is an except of a \texttt{state} file:
\begin{verbatim}
 output_stdout_sigma_equil_L_2=  T
 checkpoint_period=         2000
 M_grid_size=          100
 n_part=          384
 Lx=    20.828469299691541        22.726435154004815     
 Ly=    21.821584954680326        19.440822063915213     
 Lz=    20.313660823118358        20.897094137158106     
 V=    9232.7662932888543     
 lattice=            2
 E_1=   -2398.4613158910693     
 E_2=   -2417.6552334642806     
 E=   -2417.6552334642820     
 M=    19.193917573211365     
 macro=           87
 eta=    0.0000000000000000     
 switchscalex=    1.0911236359717214     
 switchscaley=   0.89089871814033939     
 switchscalez=    1.0287212294780348     
 sweeps=          200
 moves=       154044
 moves_lattice=        77022
 accepted_moves_lattice=            1
 moves_part=        76800
\end{verbatim}
Table \ref{table:state_variables} provides a list of simulation variables, not already covered by Table \ref{table:params_in_variables},
which can be found in \texttt{state} and could be of interest to the user.

\begin{landscape}
\begin{center}\label{table:state_variables}
\begin{longtable}{ l l p{10cm}}
\caption{Useful variables found in the \texttt{monteswitch} output file \texttt{state}.}
\\

Variable & Fortran type & Description \\
\hline
\textbf{n\_part} & \texttt{INTEGER} & Number of particles in the system. \\
\textbf{Lx} & \texttt{REAL(2)} & Dimension of supercells in x-direction: the first value pertains to phase 1 while the second pertains to phase 2.\\
\textbf{Ly} & \texttt{REAL(2)} & Dimension of supercells in y-direction: the first value pertains to phase 1 while the second pertains to phase 2.\\
\textbf{Lz} & \texttt{REAL(2)} & Dimension of supercells in z-direction: the first value pertains to phase 1 while the second pertains to phase 2.\\
\textbf{V} & \texttt{REAL} & Current volume of the system.\\
\textbf{lattice} & \texttt{INTEGER} & Current phase of the system (1 or 2).\\
\textbf{E\_1} & \texttt{REAL} & Energy of phase 1 for the current displacements. \\
\textbf{E\_2} & \texttt{REAL} & Energy of phase 2 for the current displacements. \\
\textbf{E} & \texttt{REAL} & Current energy of the system. This is \textbf{E\_1} if we are in phase 1 and \textbf{E\_2} if we are in phase 2.\\
\textbf{M} & \texttt{REAL} & Current order parameter of the system.\\
\textbf{macro} & \texttt{REAL} & Current macrostate of the system.\\
\textbf{eta} & \texttt{REAL} & The value of the weight function associated with the current macrostate of the system.\\
\textbf{switchscalex} & \texttt{REAL} & Scaling of the supercell in the x-dimension when performing a lattice switch from phase 1 to phase 2. The 
reciprocal of this is the scaling when performing a lattice switch from phase 2 to phase 1.\\
\textbf{switchscaley} & \texttt{REAL} & Scaling of the supercell in the x-dimension when performing a lattice switch from phase 1 to phase 2. The 
reciprocal of this is the scaling when performing a lattice switch from phase 2 to phase 1. \\
\textbf{switchscalez} & \texttt{REAL} & Scaling of the supercell in the x-dimension when performing a lattice switch from phase 1 to phase 2. The 
reciprocal of this is the scaling when performing a lattice switch from phase 2 to phase 1. \\
\textbf{sweeps} & \texttt{INTEGER} & Number of sweeps performed so far, including over previous simulations if we have used the \texttt{-resume} argument.\\
\textbf{moves} & \texttt{INTEGER} & Total number of moves performed so far in total, including over previous simulations if we have used the \texttt{-resume} 
argument.\\
\textbf{moves\_lattice} & \texttt{INTEGER} & Number of lattice moves performed so far, including over previous simulations if we have used the 
 \texttt{-resume} argument.\\
\textbf{accepted\_moves\_lattice} & \texttt{INTEGER} & Number of accepted lattice moves so far, including over previous simulations if we have used the 
 \texttt{-resume} argument.\\
\textbf{moves\_part} & \texttt{INTEGER} & Number of particle moves performed so far, including over previous simulations if we have used the 
 \texttt{-resume} argument.\\
\textbf{accepted\_moves\_part} & \texttt{INTEGER} & Number of accepted particle moves so far, including over previous simulations if we have used the 
 \texttt{-resume} argument.\\
\textbf{moves\_vol} &\texttt{INTEGER} & Number of volume moves performed so far, including over previous simulations if we have used the 
 \texttt{-resume} argument.\\
\textbf{accepted\_moves\_vol} & \texttt{INTEGER} & Number of accepted volume moves so far, including over previous simulations if we have used the 
 \texttt{-resume} argument.\\
\textbf{melts} & \texttt{INTEGER} & The number of times the system has melted. \\
\textbf{barrier\_macro\_low} & \texttt{INTEGER} & The macrostate number corresponding to the lowest currently allowed macrostate (relevant only when artificial dynamics is enabled). \\
\textbf{barrier\_macro\_high} & \texttt{INTEGER} & The macrostate number corresponding to the highest currently allowed macrostate (relevant only when artificial
dynamics is enabled). \\
\textbf{block\_counts} & \texttt{INTEGER} & The total number of blocks considered so far for block averaging.  \\
\textbf{equil\_DeltaF} & \texttt{REAL} & The free energy difference between the phases ($F_1-F_2$; extensive) evaluated using block averaging.\\
\textbf{sigma\_equil\_DeltaF} & \texttt{REAL} & The uncertainty in \textbf{equil\_DeltaF} evaluated using block averaging..\\
\textbf{equil\_H\_1} & \texttt{REAL} & The energy (for NVT simulations) or enthalpy (for NPT simulations) of phase 1 evaluated using block averaging.\\
\textbf{equil\_H\_2} & \texttt{REAL} & The energy (for NVT simulations) or enthalpy (for NPT simulations) of phase 2 evaluated using block averaging.\\
\textbf{sigma\_equil\_H\_1} & \texttt{REAL} & The uncertainty in \textbf{equil\_H\_1}.\\
\textbf{sigma\_equil\_H\_2} & \texttt{REAL} & The uncertainty in \textbf{equil\_H\_2}.\\
\textbf{equil\_V\_1} & \texttt{REAL} & The volume of phase 1 evaluated using block averaging.\\
\textbf{equil\_V\_2} & \texttt{REAL} & The volume of phase 2 evaluated using block averaging.\\
\textbf{sigma\_equil\_V\_1} & \texttt{REAL} &  The uncertainty in \textbf{equil\_V\_1}.\\
\textbf{sigma\_equil\_V\_2} & \texttt{REAL} &  The uncertainty in \textbf{equil\_V\_2}.\\
\textbf{R\_1} & \texttt{REAL(n\_part,3)} & The current lattice vectors for phase 1. \\
\textbf{R\_2} & \texttt{REAL(n\_part,3)} & The current lattice vectors for phase 2.\\
\textbf{u} & \texttt{REAL(n\_part,3)} & The current displacement vectors.\\
\textbf{M\_grid} & \texttt{REAL(M\_grid\_size)} & Array containing the minimum order parameter for each macrostate: macrostate $n$ corresponds to 
order parameters between \textbf{M\_grid}$(n)$ and \textbf{M\_grid}$(n+1)$.\\
\textbf{M\_counts\_1} & \texttt{INTEGER(M\_grid\_size)} & \textbf{M\_counts\_1}($n$) is the number of times macrostate $n$ has been visited while the
system was in phase 1 so far, including over previous simulations if we have used the \texttt{-resume} argument.\\
\textbf{M\_counts\_2} & \texttt{INTEGER(M\_grid\_size)} & \textbf{M\_counts\_2}($n$) is the number of times macrostate $n$ has been visited while the
system was in phase 2 so far, including over previous simulations if we have used the \texttt{-resume} argument. \\
\textbf{eta\_grid} & \texttt{REAL(M\_grid\_size)} & \textbf{eta\_grid}($n$) is the value of the weight function for macrostate $n$. \\
\textbf{trans} & \texttt{REAL(M\_grid\_size,M\_grid\_size)} & \textbf{trans}($m,n$) is the number of inferred transitions from macrostate
$m$ to macrostate $n$; it is the matrix $\mathcal{C}_{\mathcal{M}\mathcal{M}'}$ in Section \ref{sec:transition_matrix}. \\
\textbf{equil\_umsd\_1} & \texttt{REAL(n\_part)} & \textbf{equil\_umsd\_1}($n$) is the mean-squared displacement of particle $n$ from its lattice site
in phase 1, evaluated using block averaging. \\
\textbf{equil\_umsd\_2} & \texttt{REAL(n\_part)} & \textbf{equil\_umsd\_2}($n$) is the mean-squared displacement of particle $n$ from its lattice site
in phase 2, evaluated using block averaging. \\
\textbf{sigma\_equil\_umsd\_1} & \texttt{REAL(n\_part)} & \textbf{sigma\_equil\_umsd\_1}($n$) is the uncertainty in \textbf{equil\_umsd\_1}($n$). \\
\textbf{sigma\_equil\_umsd\_2} & \texttt{REAL(n\_part)} & \textbf{sigma\_equil\_umsd\_2}($n$) is the uncertainty in \textbf{equil\_umsd\_2}($n$). \\
\textbf{equil\_L\_1} & \texttt{REAL(3)} & The 1st, 2nd and 3rd values in the array \textbf{equil\_L\_1} are the mean x-, y- and z-dimensions of the
supercell in phase 1, evaluated using block averaging. \\
\textbf{equil\_L\_2} & \texttt{REAL(3)} &  The 1st, 2nd and 3rd values in the array \textbf{equil\_L\_2} are the mean x-, y- and z-dimensions of the
supercell in phase 2, evaluated using block averaging. \\
\textbf{sigma\_equil\_L\_1} & \texttt{REAL(3)} & \textbf{sigma\_equil\_L\_1}($n$) is the uncertainty in \textbf{equil\_L\_1}($n$). \\
\textbf{sigma\_equil\_L\_2} & \texttt{REAL(3)} & \textbf{sigma\_equil\_L\_2}($n$) is the uncertainty in \textbf{equil\_L\_2}($n$). \\
\end{longtable}
\end{center}
\end{landscape}

\subsection{Resuming a checkpointed simulation}
The command-line argument \texttt{-resume} continues an `old' simulation, whose variables are contained in the file \texttt{state} in the current 
directory. The `resumed' simulation is run for the number of Monte Carlo sweeps specified in the variable \textbf{stop\_sweeps} in \texttt{state}.
By default this is the number of sweeps which were performed in the old simulation, though one of course this can be manually altered if one wants
the resumed simulation to be of a different length to the old simulation.
For a simulation invoked using the argument \texttt{-resume}, the file \texttt{data} is amended: the resumed simulation does not overwrite the
\texttt{data} file; all information from the old simulation is retained in it.

The command-line argument \texttt{-reset} invokes a simulation from an old \texttt{state} file similarly to \texttt{-resume}, except that it
resets all `counter variables' to zero. This has the effect of starting a `new' simulation whose nature corresponds to the old simulation, 
but instead uses the state of the system specified in \texttt{state}. By contrast, the argument \texttt{-new} initialises the 
state to be such that the particles form a perfect crystal lattice, which usually does not correspond to an equilibrated state.
By `counter variables' we mean those such as variables describing the number of moves performed for each move type, the number of accepted moves 
for each move type, and variables pertaining to equilibrium quantities.
For a simulation invoked using the argument \texttt{-reset}, the file \texttt{data} is overwritten, i.e., the information from the `old'
simulation is not retained.

\subsection{MPI simulations}
As mentioned at the beginning of this chapter, the program \texttt{monteswitch\_mpi} is the MPI-parallelised analogue of \texttt{monteswitch}. 
To build upon what was already said, \texttt{monteswitch\_mpi} is identical to the program, except that instead of a single simulation for 
\textbf{stop\_sweeps} Monte Carlo sweeps, 
\texttt{monteswitch\_mpi} runs $n$ simulations -- replicas -- in parallel using MPI, each being approximately \textbf{stop\_sweeps}$/n$ sweeps in 
length. Accordingly, during the simulation multiple \texttt{data}- and \texttt{state}-format files are created -- one for each replica. These 
are named \texttt{state\_0}, \texttt{state\_1}, \texttt{state\_2}, etc., and similarly for the \texttt{data}-format files. At the completion of 
all replicas, the results of block averaging from all replicas are combined and stored in the file \texttt{state}. The state stored in \texttt{state}
corresponds to the state in \texttt{state\_0}. 

Further details of the parallelisation are as follows. All replicas are always initialised to be in the same state. For a new simulation this is 
determined by the \textbf{init\_lattice} variable in \texttt{params\_in} similarly to \texttt{monteswitch}. For a resumed simulation this is the state 
contained in the \texttt{state} file from which the simulation is to be resumed. We emphasise that \emph{all replicas of the system are initialised with 
the same state} when the \texttt{-resume} argument is used with \texttt{monteswitch\_mpi}. 
The files \texttt{data\_}$n$ are therefore always overwritten by \texttt{monteswitch\_mpi} when the \texttt{-resume} argument is
invoked since, given the nature of the parallelisation, there is no continuity between the replicas in subsequent simulations. The exception is replica `0',
whose state is always stored in the file \texttt{state}, as well as \texttt{state\_0}.

\section{Utility programs}\label{sec:utility_programs}
As mentioned earlier, \emph{monteswitch} contains a number of utility programs which assist with post-processing of the data and the construction
of input files. We now describe these programs.

\subsection{Programs for generating \texttt{lattices\_in} files}
The programs \texttt{lattices\_in\_hcp\_fcc}, \texttt{lattices\_in\_bcc\_fcc} and \texttt{lattices\_in\_bcc\_hcp} 
create \texttt{lattices\_in} files containing reference states corresponding to, respectively, hcp--fcc, bcc--fcc and bcc--hcp lattice switches.
Usage of these programs is as follows:
\begin{verbatim}
lattices_in_hcp_fcc <rho> <nx> <ny> <nz>
lattices_in_bcc_fcc <rho> <nx> <ny> <nz>
lattices_in_bcc_hcp <rho> <nx> <ny> <nz>
\end{verbatim}
The command-line arguments \texttt{<rho>}, \texttt{<nx>}, \texttt{<ny>} and \texttt{<nz>} constitute the free parameters for the 
\texttt{lattices\_in} file to be created.
The first argument \texttt{<rho>} is the density (i.e., the number of particles per unit volume) of the reference states to
construct. (Note that both states necessarily have the same density). The second, third and fourth arguments \texttt{<nx>}, 
\texttt{<ny>} and \texttt{<nz>} are integers which correspond to the number of unit cells (described in a moment) which will 
be tiled in the x-, y- and z-directions respectively to construct the supercell for each phase. 
The programs output the desired \texttt{lattices\_in} file to stdout. Hence one must redirect the output
to create the required \texttt{lattices\_in} file, e.g., \verb|$ ./lattices_in_hcp_fcc 0.5 2 3 5 > lattices_in|

What follows is a description of the unit cells for each pair of phases for each program. 

\subsubsection{\texttt{lattices\_in\_hcp\_fcc}}
The unit cell here contains 12 particles. The particles are spread over 6 planes in the z-direction; each plane contains two particles. The positions
of the particles corresponds to a stacking sequence for the planes of ABCABC for the fcc unit cell, and  ABABAB for the hcp unit cell. 

\subsubsection{\texttt{lattices\_in\_bcc\_fcc}}
The unit cell here is the conventional 2-particle body-centred tetragonal (bct) unit cell; for the bcc(fcc) lattice the relative dimensions of 
the bct unit cell in each Cartesian direction correspond to the bct representation of the bcc(fcc) lattice.

\subsubsection{\texttt{lattices\_in\_bcc\_hcp}}
The unit cell here contains 4 particles. The bcc unit cell is the 4-particle face-centred tetragonal (fct) corresponding to the fct representation of 
the bcc lattice. The hcp unit cell is the `fct-like' representation of the hcp lattice.

\subsection{Post-processing \texttt{state} files}\label{sec:monteswitch_post}
\texttt{monteswitch\_post} is a tool for post-processing the file \texttt{state} generated by \texttt{monteswitch} or \texttt{monteswitch\_mpi}. It can
be used to extract useful information from that file. The \texttt{state} file which the program operates on is that in the current directory.
The command-line arguments determine the task performed by the program. Usage of \texttt{monteswitch\_post} is as follows, where the function 
of each command-line argument is described below:
\begin{verbatim}
monteswitch_post -extract_wf
monteswitch_post -extract_M_counts
monteswitch_post -extract_pos [<species>]
monteswitch_post -extract_R_1 [<species>]
monteswitch_post -extract_R_2 [<species>]
monteswitch_post -extract_u [<species> <phase>]
monteswitch_post -calc_rad_dist <bins>
monteswitch_post -merge_trans <state_in_1> <state_in_2> <state_out>
monteswitch_post -extract_lattices_in <vectors_in_1> <vectors_in_2>
monteswitch_post -extract_pos_xyz
monteswitch_post -set_wf <wf_file>
monteswitch_post -taper_wf
\end{verbatim}

\subsubsection{\texttt{-extract\_wf}}
Extract the weight function from \texttt{state} and output it to stdout. In the output the first token on each line is the order parameter, 
and the second is the corresponding value of the weight function.

\subsubsection{\texttt{-extract\_M\_counts}}
Extract order parameter histograms from \texttt{state} and output them to stdout. In the output the first token on each line is the order 
parameter, the second is the corresponding number of counts for phase 1, and the third is the corresponding number of counts for 
phase 2. 

\subsubsection{\texttt{-extract\_pos [<species>]}}
Extract the current positions of the particles, and output them to stdout. In the output the first, second and third tokens on each line 
are the x-, y- and z-coordinates respectively for a particle. If the optional argument \texttt{<species>} is present then only the positions
for particles belonging to species \texttt{<species>} are output.

\subsubsection{\texttt{-extract\_R\_1 [<species>]}}
Extract the current positions of the lattice sites for phase 1, and output them to stdout. In the output the first, second and third 
tokens on each line are the x-, y- and z-coordinates respectively for a particle. If the optional argument \texttt{<species>} is present 
then only the sites for particles belonging to species \texttt{<species>} in phase 1 are output.

\subsubsection{\texttt{-extract\_R\_2 [<species>]}}
Extract the current positions of the lattice sites for phase 2, and output them to stdout. In the output the first, second and third 
tokens on each line are the x-, y- and z-coordinates respectively for a particle. If the optional argument \texttt{<species>} is present 
then only the sites for particles belonging to species \texttt{<species>} in phase 2 are output.

\subsubsection{\texttt{-extract\_u [<species> <phase>]}}
Extract the displacements of the particles, and output them to stdout. In the output the first, second and third tokens on each line are 
the x-, y- and z-displacements respectively for a particle. If the optional arguments \texttt{<species>} and \texttt{<phase>} are present 
then only the displacements for particles belonging to species \texttt{<species>} in phase \texttt{<phase>} are output.

\subsubsection{\texttt{-calc\_rad\_dist <bins>}}
Calculate the radial distribution function, based on the current state of the system, and output it to stdout. \texttt{<bins>} is the number of bins to 
be used in the function.

\subsubsection{\texttt{-merge\_trans} \texttt{<state\_in\_1>}  \texttt{<state\_in\_2>} \texttt{<state\_out>}}
Combine the \textbf{trans} matrices from the files \texttt{<state\_in\_1>} and \texttt{<state\_in\_2>}, and store the result in the 
file \texttt{<state\_out>}, where all variables in \texttt{<state\_out>} other than the matrix \textbf{trans} are inherited from 
\texttt{<state\_in\_1>}. This argument can be used for pooling the results of multiple simulations which utilise the same underlying
`order parameter grid' \textbf{M\_grid\_size}.

\subsubsection{\texttt{-extract\_lattices\_in <vectors\_in\_1> <vectors\_in\_2>}}
Output the geometrical properties of the system in the format of a \texttt{lattices\_in} file. The arguments \texttt{<vectors\_in\_1>}
and \texttt{<vectors\_in\_2>} can be either \texttt{pos} or \texttt{R}. If \texttt{<vectors\_in\_1>} is \texttt{pos}, then
the positions of the particles in phase 1 of the forthcoming \texttt{lattices\_in} file will be the positions of the particles in phase 1
in the \texttt{state} file; if \texttt{<vectors\_in\_1>} is \texttt{R}, then the positions of the particles in phase 1 of the 
\texttt{lattices\_in} file will be the current lattice vectors (i.e., \textbf{R\_1}) corresponding to phase 1 in the \texttt{state} file.
Similar applies for \texttt{<vectors\_in\_2>} with phase 2.

\subsubsection{\texttt{-extract\_pos\_xyz}}
Extract the positions of the particles, and output them to stdout in `.xyz' format. In the output the first line contains the number of 
particles, the second line is a comment line, and the subsequent lines contain the particle positions and species: the first token is 
the `element' (set to `A' for species `1', `B' for species `2', ..., `Z' for species 26, and `?' otherwise), and the second, third, 
fourth and fifth tokens are the x-, y- and z-coordinates respectively.

\subsubsection{\texttt{-set\_wf <wf\_file>}}
Alters the weight function in \texttt{state} to correspond to that specified in the file \texttt{<wf\_file>}. The format of the file
\texttt{<wf\_file>} must be analogous to the format of the weight function output by this program via the \texttt{-extract\_wf} argument
described above: the file must contain \textbf{M\_grid\_size} lines, each containing two tokens (extra lines and tokens are ignored),
which both should be of type \texttt{REAL}. The first token on each line is ignored, while the second tokens are treated as the new
weight function: the value of the weight function for macrostate $i$ is set to the value of the second token on line $i$ in
\texttt{<wf\_file>}.

\subsubsection{\texttt{-taper\_wf}}\label{sec:taper}
`Tapers' the weight function in the \texttt{state} file. To elaborate, it is assumed that the weight function has a single local 
minimum in the $M<0$ region (the region associated with phase 1) and a single local minimum in the $M>0$ region (associated with phase 2), with 
a maximum at $M\approx 0$ separating the two minima. Let $\eta_{\text{min},1}$ denote the value of the weight function at the minimum in the $M<0$
region, and let $M_{\text{min},1}$ denote the location of this minimum. In the $M<0$ region the weight function is tapered by setting
$\eta_M=\eta_{\text{min},1}$ for all $M<M_{\text{min},1}$. Similarly the weight function is tapered in the $M>0$ region by setting 
$\eta_M=\eta_{\text{min},2}$ for all $M>M_{\text{min},2}$. Thus the weight function is `flattened' for the regions in $M$-space `outwith' the two minima. 
This prevents oversampling of these regions, which is unnecessary and inefficient.

\section{Examples}\label{sec:example}
We now present results obtained using \emph{monteswitch} for two systems in order to elucidate how \emph{monteswitch} could be used in practice. 

\subsection{Hard-sphere solid: hcp vs. fcc}
The first system we consider is the hard-sphere solid. This system has been studied extensively with LSMC 
\cite{Bruce_1997,Pronk_1999,Bruce_2000,Bridgwater_2014}, and accordingly makes
an excellent testing ground for \emph{monteswitch}. LSMC studies of this system have focused on calculating the free energy difference between
the hcp and fcc phases. We have done the same using \emph{monteswitch} for a 216-particle system, with the aim of validating \emph{monteswitch}
against the results of other studies. Specifically we consider the NVT ensemble at reduced density $\tilde{\rho}=0.7778$,
where $\tilde{\rho}$ is related to the number density of the system $\rho$ by the equation $\tilde{\rho}=\rho/\sqrt{2}$; and the NPT ensemble at
$P\beta\sigma^3=14.58$, where $\sigma$ denotes the hard sphere diameter.
For the NPT ensemble we consider both \emph{anisotropic} volume moves which allow the Cartesian dimensions of the supercell to vary independently, 
and \emph{isotropic} volume moves in which the supercell's shape, but not size, is constrained.
Note that for each of the three ensembles we considered,  i.e., the NVT ensemble, the isotropic NPT ensemble and the anisotropic NPT ensemble,
we used the utility program \texttt{lattices\_in\_hcp\_fcc} (see Section \ref{sec:utility_programs}) 
to generate the relevant \texttt{lattices\_in} input file required by the main programs \texttt{monteswitch} and \texttt{monteswitch\_mpi}
(Section \ref{sec:simulation_programs}).

\begin{figure}
\centering
\includegraphics[height=\textwidth,angle=270]{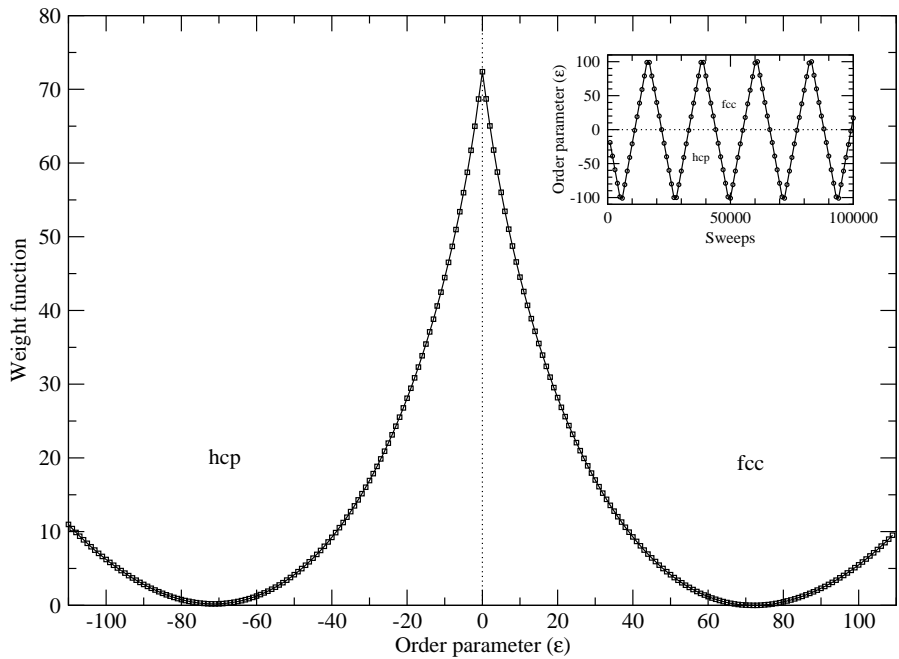}
\caption
{Weight function for an LSMC simulation of a 216-particle hard-sphere system in the NVT ensemble at $\tilde{\rho}=0.7778$; and the evolution of 
the order parameter vs. simulation time for one of the replicas during the corresponding weight-function-generation simulation 
(inset; only the first 100,000 sweeps are shown). 
The regions of order-parameter space corresponding to hcp and fcc are indicated. The order parameter is in units of the sphere overlap energy
$\epsilon$, which is set sufficiently high during the simulation to realise hard spheres (see Section \ref{sec:penetrable_spheres}).}
\label{fig:HS_generation_results}
\end{figure}

For each of the three ensembles we used the following procedure to obtain the hcp--fcc free energy difference, and, in the case of the NPT ensembles,
the densities of each phase $\tilde{\rho}_{\text{hcp}}$ and $\tilde{\rho}_{\text{fcc}}$.
We first performed a number of short conventional Monte Carlo simulations in each phase. The aim of these was to determine quantities to be used in
the forthcoming LSMC simulations, such as the maximum particle and volume move sizes, and the appropriate range of order-parameter space to consider. 
Then we performed an LSMC simulation in order to generate the weight function. In the weight-function-generation simulations we used artificial 
dynamics to quickly generate an accurate weight function. Moreover we used \texttt{monteswitch\_mpi} to parallelise the simulation: we 
ran 16 replicas of the system in parallel. Our weight-function-generation simulations each consisted of 18,000,000 Monte Carlo sweeps (1,125,000 
sweeps per replica). 
Fig. \ref{fig:HS_generation_results} shows the weight function obtained from a weight-function-generation simulation for the NVT ensemble, 
along with the evolution of the order parameter for one of the replicas during that simulation. Note that order-parameter space is explored systematically, 
concordant with artificial dynamics as described in Section \ref{sec:exploring_M_space}.
Using this weight function we then we performed `production' LSMC simulations to determine the hcp--fcc free energy difference -- and also
$\tilde{\rho}_{\text{hcp}}$ and $\tilde{\rho}_{\text{fcc}}$ for the NPT systems. For each ensemble we
performed two production simulations, where each production simulation consisted of 125,000,000 sweeps, again using 16 replicas in parallel via 
\texttt{monteswich\_mpi} (7,812,500 sweeps per replica). Thus the total number of production sweeps for each ensemble was 250,000,000.
Finally, the results from all production replicas for each ensemble were pooled to obtain the final results.

The final results are presented in Table \ref{table:HS_results} along with the results of other studies. As can be seen from the table \emph{monteswitch} 
is in agreement with the other studies.
Note that slightly different NPT results are obtained using isotropic and anisotropic volume moves. This is expected since
isotropic moves constrain the underlying lattice for each phase to be `strictly' hcp or fcc, while anisotropic moves allow the system to `stretch' 
along any Cartesian dimension.
Interestingly the densities of both phases are higher when isotropic moves are used.

\begin{table}\label{table:HS_results}
\begin{center}
\begin{tabular}{lllll}
System            & Study & $\beta\Delta\mathcal{F}_{\text{fcc$\to$hcp}}$ & $\tilde{\rho}_{\text{hcp}}$ & $\tilde{\rho}_{\text{fcc}}$ \\
                                                        \hline
NVT, $\tilde{\rho}=0.7778$ & \emph{This work}             & 0.00135(5) & -                        & -                        \\
                            & Ref. \cite{Pronk_1999}      & 0.00132(4) & -                        & -                        \\
                            & Ref. \cite{Bruce_2000}      & 0.00133(4) & -                        & -                        \\
                            & Ref. \cite{Bridgwater_2014} & 0.00133(3) & -                        & -                        \\

NPT, $P\beta\sigma^3=14.58$, iso & \emph{This work}       & 0.00123(6) & 0.77820(6)               & 0.77820(6)               \\

NPT, $P\beta\sigma^3=14.58$, aniso & \emph{This work}     & 0.00117(9) & 0.77759(6)               & 0.77768(6)               \\
                            & Ref. \cite{Bruce_2000}      & 0.00113(4) & 0.7776(1)                & 0.7775(1)                \\

\end{tabular}
\caption{LSMC results for various 216-particle hard-sphere systems. $\Delta\mathcal{F}_{\text{fcc$\to$hcp}}$ is the \emph{intensive}
Helmholtz(Gibbs) free energy difference in the NVT(NPT) ensemble. `iso' refers to isotropic volume moves, while `aniso' refers to anisotropic
volume moves in which the shape of the supercell is allowed to vary independently in each Cartesian dimension.
Uncertainties in the LSMC quantities of this work are standard errors of the mean.}
\end{center}
\end{table}

\subsection{Embedded atom model for Zr: bcc vs. hcp}
As our second example we consider the hcp--bcc transition in Zr, modeled using the EAM potential `\#2' 
developed in Ref. \cite{Mendelev_2007}. 
In Ref. \cite{Mendelev_2007} the authors determined that the zero-pressure transition temperature $T_{\text{hcp$\to$bcc}}$ for this potential
was 1233K. They also determined that the enthalpy change $\Delta H_{\text{hcp$\to$bcc}}$ and fractional volume change $\Delta V_{\text{hcp$\to$bcc}}/V_{\text{hcp}}$ 
associated with the transition were 0.039 eV/atom and -0.8\% respectively.
However these quantities were determined \emph{indirectly} from the results of multiple molecular dynamics simulations 
(see Ref. \cite{Mendelev_2007} for details). 
By contrast we have used \emph{monteswitch} to determine $T_{\text{hcp$\to$bcc}}$, $\Delta H_{\text{hcp$\to$bcc}}$ and $\Delta V_{\text{hcp$\to$bcc}}/V_{\text{hcp}}$ 
\emph{directly} via LSMC. 
Note that this is a more `realistic' example than the previous one: usually one is interested in, e.g., the temperature
and pressure at which a phase transition between two phases occurs, as opposed to the value of the free energy difference between the phases at one
state point (which was the nature of the previous example).

While LSMC provides a means for calculating the Gibbs free energy difference $\Delta G$ between two phases at a given 
temperature and pressure, there is the question of how to exploit LSMC to determine the transition temperature, which by definition is the temperature 
at which $\Delta G=0$. We used the following iterative procedure, which in Ref. \cite{Jackson_2002} was shown, for the Lennard-Jones solid,
to outperform other existing methods.
We first considered an initial estimate for the transition temperature $T^{(1)}$. We then used LSMC to obtain the Gibbs free energy difference between 
the phases $\Delta G\equiv G_{\text{bcc}}-G_{\text{hcp}}$, the enthalpies and volumes for each phase ($H_{\text{hcp}}$, $H_{\text{bcc}}$, $V_{\text{hcp}}$ and 
$V_{\text{bcc}}$), as well as uncertainties in these quantities, at $T^{(1)}$. We then substituted these quantities into the following equation to generate 
a more accurate estimate for the transition temperature $T^{(2)}$:
\begin{equation}\label{Newton-Raphson}
T^{(n+1)}=T^{(n)}\Biggl[1-\Biggl(1-\frac{\Delta H^{(n)}}{\Delta G^{(n)}}\Biggr)\Biggr],
\end{equation}
where $\Delta H\equiv H_{\text{bcc}}-H_{\text{hcp}}$, and the superscript `$(n)$' signifies a quantity obtained from an LSMC simulation at temperature $T^{(n)}$.
(The above equation is, in fact, the Newton-Raphson estimate of the temperature $T^{(n+1)}$ at which $\Delta G^{(n+1)}=0$; see Ref. \cite{thesis:Jackson}
for details).
We then repeated all of the above for $T^{(2)}$, $T^{(3)}$, etc. until the procedure converged. The procedure was deemed to have converged if the change 
in temperature for the next iteration, $T^{(n+1)}-T^{(n)}$, was less than its corresponding uncertainty -- 
calculated from propagating the uncertainties in $\Delta H^{(n)}$ and $\Delta G^{(n)}$ through Eqn. \eqref{Newton-Raphson}. 
At this point the next iteration was not performed, and the enthalpies and volumes obtained from iteration $n$ were used to form our quoted values for
$\Delta H_{\text{hcp$\to$bcc}}$ and $\Delta V_{\text{hcp$\to$bcc}}/V_{\text{hcp}}$; and $T^{(n+1)}$ was used as our quoted value for $T_{\text{hcp$\to$bcc}}$.

\begin{figure}
\centering
\includegraphics[height=\textwidth,angle=270]{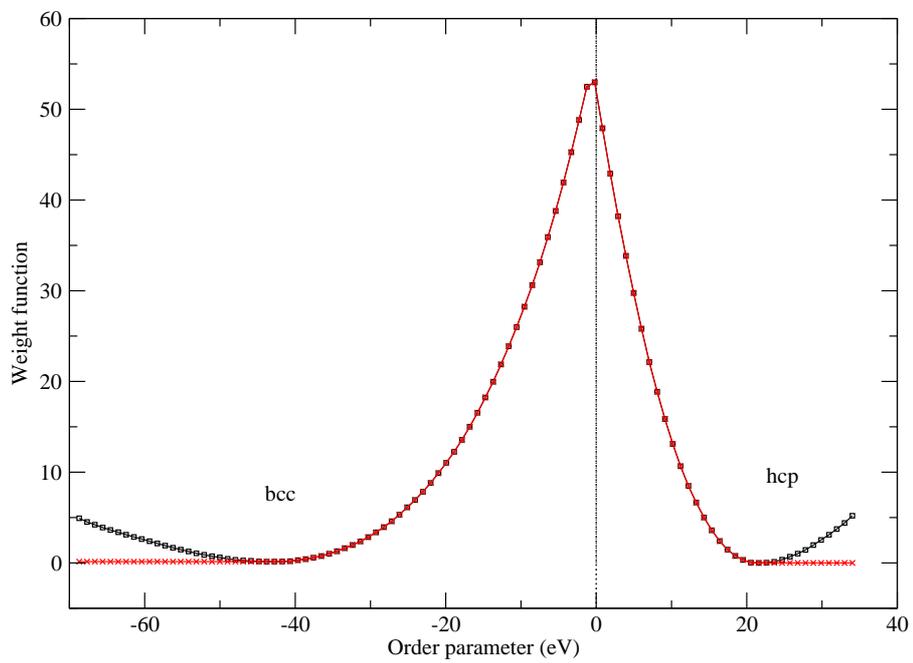}
\caption
{Weight function for Zr potential `\#2' of Ref. \cite{Mendelev_2007} applied to a 384-atom system at $T=1275.86$K. 
The black curve with squares corresponds to the `raw' weight function obtained from the weight-function-generation simulation, while the red curve 
with crosses corresponds to the weight function after `tapering'. The regions of order-parameter space corresponding to bcc and hcp are indicated.}
\label{fig:EAM_weight_function}
\end{figure}

We considered two system sizes: a smaller system containing 384 atoms and a larger system containing 1296 atoms.
Note that both these systems are significantly smaller than those used in Ref. \cite{Mendelev_2007}. For both system sizes we used the NPT ensemble with 
anisotropic volume moves (in the sense described in the previous example), and bootstrapped the iterative procedure with $T^{(1)}$=900K. Furthermore
we used the utility program \texttt{lattices\_in\_bcc\_hcp} to generate the relevant \texttt{lattices\_in} input files for all simulation.
Our procedure for calculating $\Delta G$, $H_{\text{hcp}}$, $H_{\text{bcc}}$, $V_{\text{hcp}}$ and $V_{\text{bcc}}$ at a given temperature is similar to that
described in the previous example: we first performed short conventional Monte Carlo simulations in each phase, then a weight-function-generation simulation,
and finally a production simulation. 
Noteworthy aspects of the procedure are as follows. Firstly, our weight-function-generation simulations consisted 
of 160,000 Monte Carlo sweeps using 4 replicas in parallel via \texttt{monteswitch\_mpi} (40,000 sweeps per replica).
Secondly, we `tapered' the weight function obtained from a weight-function-generation simulation using the 
utility program \texttt{monteswitch\_post} before using the weight function in a production simulation.
Fig. \ref{fig:EAM_weight_function} shows the weight function obtained for the 384-atom system at $T=1275.86$K (the final
iteration, $n=3$, as is discussed below) before and after tapering.
Tapering was done in order to improve the efficiency of the production simulation (see Section \ref{sec:taper} for details). 
Finally, at each temperature we performed a single production simulation consisting of 700,000 sweeps, again using 4 replicas in 
parallel (175,000 sweeps per replica).

Our results are presented in Table \ref{table:EAM_production_results}, where they are compared against those of Ref. \cite{Mendelev_2007}. 
Note that our 384-atom and 1296-atom results are in agreement with each other, which means that finite-size effects are insignificant
in the 384-atom system for the quantities we consider here. Note also that the iterative procedure converged
quickly for both system sizes: only a few iterations were needed to pinpoint the transition temperature to an accuracy of $\lesssim$ 2K.
We emphasise that only relatively modest computational effort was required to obtain this accuracy: for the 384-atom system each iteration 
(i.e., one 160,000-sweep weight-function-generation simulation and one 700,000-sweep production simulation) took a wall-clock time of 
$\approx$1.9 hours on a desktop machine (exploiting 4 cores),
\footnote{Specifically, the machine we used was an iMac14,2 with a 3.2GHz Intel Core i5-4570 processor.}
while for the 1296-atom system each iteration took $\approx$ 17.7 hours.
Our results are in agreement with those of Ref. \cite{Mendelev_2007} for $\Delta H_{\text{hcp$\to$bcc}}$ and $\Delta V_{\text{hcp$\to$bcc}}/V_{\text{hcp}}$. 
Regarding $T_{\text{hcp$\to$bcc}}$, we find $T_{\text{hcp$\to$bcc}}$ to be 42K higher than the 1233K quoted in Ref. \cite{Mendelev_2007}. 
Whether this is sufficiently close to 1233K to constitute `agreement' depends on the uncertainty in the Ref. \cite{Mendelev_2007} value, 
which unfortunately is not provided in that study. However in Ref. \cite{Sun_2006} the same method as in Ref. \cite{Mendelev_2007}
was used to determine $T_{\text{hcp$\to$bcc}}$ for a Mg EAM potential; moreover the uncertainty in the value was stated to be 40K. 
(Specifically, $T_{\text{hcp$\to$bcc}}$ was found to be 645$\pm$40K for the Mg potential). If this uncertainty carries over to the Zr study of 
Ref. \cite{Mendelev_2007}, then it follows that our $T_{\text{hcp$\to$bcc}}$ is in agreement with that of Ref. \cite{Mendelev_2007}.

\begin{table}\label{table:EAM_production_results}
\begin{center}
\begin{tabular}{l l l l l l l}
         & $n$ & $T^{(n)}$ & $\Delta G_{\text{hcp$\to$bcc}}$ & $\Delta H_{\text{hcp$\to$bcc}}$  & $\Delta V_{\text{hcp$\to$bcc}}/V_{\text{hcp}}$ & $T_{\text{hcp$\to$bcc}}$ \\
\hline
$N=384$  &  1  & \phantom{0}900.00  & \phantom{-}0.0114(11)   & 0.0459(4)       & -0.8(2)        & 1200(40)      \\
         &  2  & 1199.91            & \phantom{-}0.00238(3)   & 0.0401(2)       & -0.810(6)      & 1275.9(11)    \\
         &  3  & 1275.86            & -0.00003(5)             & \bf{0.0379(2)}  & \bf{-0.787(7)} & \bf{1275(2)}  \\

$N=1296$ &  1  & \phantom{0}900.00  & \phantom{-}0.01265(2)   & 0.04589(7)      & -0.996(3)      & 1242.6(11)    \\
         &  2  & 1242.61            & \phantom{-}0.001031(10) & 0.03930(6)      & -0.791(7)      & 1276.1(3)     \\
         &  3  & 1276.10            & -0.00003(2)             & 0.03835(13)     & -0.772(5)      & 1275.1(6)     \\
         &  4  & 1275.13            & -0.000000(10)           & \bf{0.0383(2)}  & \bf{-0.772(7)} & \bf{1275.1(3)} \\

Ref. \cite{Mendelev_2007} & - & -   & -                       & 0.039           & -0.8           & 1233           \\

\end{tabular}
\caption{Results of LSMC simulations for Zr potential `\#2' of Ref. \cite{Mendelev_2007}. Results are given for each iteration $n$ of the iterative
procedure described in the main text, for each system size considered ($N$ denotes the number of atoms in the system). The results of the final iteration,
which are to be compared to the results of Ref. \cite{Mendelev_2007} (bottom row), are in bold text. Note that for the LSMC results 
$T_{\text{hcp$\to$bcc}}$ is the transition temperature as predicted from the results of the corresponding LSMC simulation, whereas $T^{(n)}$ 
is the temperature at which that simulation was performed.
All temperatures are in K, $\Delta G_{\text{hcp$\to$bcc}}$ and $\Delta H_{\text{hcp$\to$bcc}}$ are in eV/atom, and $\Delta V_{\text{hcp$\to$bcc}}/V_{\text{hcp}}$ is
a percentage. Uncertainties in the LSMC quantities are standard errors of the mean.}
\end{center}
\end{table}

\section{Conclusions and outlook}\label{sec:conclusions}
We have described \emph{monteswitch}, a package for performing lattice-switch Monte Carlo (LSMC) simulations for atomic systems, and have presented
results demonstrating its efficacy.
While here we have only presented results for single-component crystalline phases, we emphasise that \emph{monteswitch} 
can be applied more generally. \emph{monteswitch} could be used to calculate the free energy difference between two multicomponent 
phases. Moreover \emph{monteswitch} could be used to evaluate the free energy of an interface between two solids, or the free energy
of a crystallographic defect. 
The free energy of an interface between two phases $A$ and $B$ can be calculated via LSMC by choosing the two supercells used in the LSMC 
simulation to have the same amounts of $A$ and $B$, but different amounts of $A$--$B$ interface \cite{Pronk_1999}. 
For example if the first supercell were comprised of superimposed $A$ and $B$ slabs with a stacking sequence $A|B|A|B$, and similarly for the 
second supercell but with a stacking sequence $A|A|B|B$, then, while both supercells contain two $A$ and $B$ slabs, the first supercell contains 
four $A$-$B$ interfaces (taking into account periodic boundary conditions) but the second contains only two interfaces. 
Denoting the area associated with one interface in the 
supercell as $\mathcal{A}$, it follows that calculating the free energy difference between these supercells via LSMC would yield the free energy 
difference associated with an area $2\mathcal{A}$ of an $A$--$B$ interface -- assuming that the slabs are sufficiently large that interfaces do not 
`interact' with one another. A similar approach could conceivably be applied to calculate the free energy of planar defects such as twin boundaries.
In principle point defects are also accessible to LSMC. For example the free energy of a Frenkel defect
could be evaluated by having one supercell be the `defect-free' crystal, while the other contains a single Frenkel defect. 
The prospect of using LSMC in this way requires further exploration.

The main strength of \emph{monteswitch} is its versatility regarding the interatomic potentials it can implement, which we have demonstrated here 
by using \emph{monteswitch} in conjunction with an embedded atom model (which is a many-body potential), and the hard-sphere model.
This versatility is achieved by having the source code for the potentials housed within a Fortran module which is amenable to customisation. As well as
using modules included with the package which implement some commonly-used potentials, it is anticipated that users will wish to write their own 
versions of this module to implement their own potentials. Templates and guidance are provided with the package to facilitate this.
An especially interesting prospect is to develop modules which interface \emph{monteswitch} with quantum chemistry programs, in order to
calculate the energy using \emph{ab initio} methods. Such `\emph{ab initio} LSMC' would be a valuable tool in examining the phase stability of
systems in which classical models are inappropriate.
Needless to say any new modules we develop will be made available to the wider community on the home page for the package \cite{website:monteswitch}.

\section*{Acknowledgements}
This work was supported by the Engineering and Physical Sciences Research Council [a Doctoral Prize Fellowship; and grant EP/M011291/1]. 
Valuable discussions with Mikhail Mendelev, Nigel Wilding, Andrey Brukhno and Kevin Stratford are gratefully acknowledged. This work made 
use of the Balena High Performance Computing Service at the University of Bath.









\bibliographystyle{elsarticle-num} 
\bibliography{bibliography}

\end{document}